\PassOptionsToPackage{hyphens}{url}

\documentclass[conference]{IEEEtran}

\usepackage{stfloats}

\usepackage{tikz}
\usepackage{amsmath}
\usepackage{algorithmic}
\usepackage{algorithm}
\usepackage{graphicx}
\usepackage{textcomp}
\usepackage{xcolor}
\usepackage{xspace}
\usepackage{enumitem}
\usepackage{caption}
\usepackage[T1]{fontenc}
\usepackage[english]{babel}
\usepackage[utf8]{inputenc}
\usepackage{longtable}
\usepackage{multirow, makecell}
\usepackage{tikz}
\usepackage{wrapfig}
\usepackage{pifont}
\usepackage{url}
\usepackage{balance}
\usepackage{afterpage}
\usepackage{multicol}
\usepackage{hyperref}
\usepackage[normalem]{ulem} 

\newsavebox{\shortpagebox}

\makeatletter
\newcommand{\shortpage}[1]
{\par
	\setbox\shortpagebox=\vbox{\strut #1\par}%
	\afterpage{\onecolumn
		\begin{multicols}{2}
			\unvbox\AP@partial
	\end{multicols}}%
	\unvbox\shortpagebox
	\par}
\makeatother

\usepackage{filecontents}
\usepackage{epstopdf}
\DeclareGraphicsExtensions{.eps}

\begin{document}
	
	\date{}

	\newenvironment{Itemize}%
	{\begin{itemize}%
			\setlength{\itemsep}{1pt}%
			\setlength{\leftmargin}{-20pt}
			\setlength{\topsep}{1pt}%
			\setlength{\partopsep}{0pt}%
			\setlength{\parskip}{0pt}}%
		{\end{itemize}}
	
	\newcommand{\namenospace}{Peekaboo}
	\newcommand{\name}{Peekaboo\xspace}
	
	\newcommand{\haojian}[1]{\textcolor{violet}{#1}}
	\newcommand{\YA}[1]{\textcolor{green}{YA: #1}} 
	\newcommand{\jason}[1]{\textcolor{blue}{[Jason: #1]}}
	\newcommand{\swarun}[1]{\textcolor{blue}{#1}}
	\newcommand{\new}[1]{\textcolor{black}{#1}}
	\newcommand{\final}[1]{\textcolor{black}{#1}}
	
		
\title{\Large \bf Peekaboo: A Hub-Based Approach to Enable Transparency in Data Processing within Smart Homes (Extended Technical Report)} 
		


\author{
	Haojian Jin, Gram Liu, David Hwang, Swarun Kumar, Yuvraj Agarwal, Jason I. Hong\\
	Carnegie Mellon University
}

\maketitle
\pagestyle{plain}

\begin{abstract}

We present \name\footnote{This technical report is an extended version of a IEEE S\&P 2022 paper~\cite{peekaboo-oakland}. We include a preliminary developer study to evaluate \name's usability and more comprehensive discussions, which were skipped due to space constraints.}, a new privacy-sensitive architecture for smart homes that leverages an in-home hub to pre-process and minimize outgoing data in a structured and enforceable manner \textit{before} sending it to external cloud servers.
\name's key innovations are (1) abstracting common data pre-processing functionality into a small and fixed set of chainable operators, and (2) requiring that developers explicitly declare desired data collection behaviors (e.g., data granularity, destinations, conditions) in an application manifest, which also specifies how the operators are chained together. Given a manifest, \name assembles and executes a pre-processing pipeline using operators pre-loaded on the hub.
In doing so, developers can collect smart home data on a need-to-know basis; third-party auditors can verify data collection behaviors; and the hub itself can offer a number of centralized privacy features to users across apps and devices, without additional effort from app developers. 
We present the design and implementation of \name, along with an evaluation of its coverage of smart home scenarios, system performance, data minimization, and example \new{built-in} privacy features.



\end{abstract}


\section{Introduction}

For many smart home products - such as smart speakers, cameras, and thermostats
- the ``brains'' of these systems are typically in the cloud. However, a key concern with cloud-based software architectures is data privacy~\cite{wang2019riverbed}. It is challenging for users to have any assurances of privacy \textit{after} their sensitive data leaves the confines of their home~\cite{Thousand7:online}. Further, it is challenging for companies to \textit{legitimately avoid collecting unnecessary smart home data while also re-assuring users and independent auditors that this is indeed the case}. 

For example, imagine a developer of a smart TV claims to only send aggregated viewing history data to their servers once a week. \textbf{How can outsiders validate this claim, since the hardware, firmware, and backend servers are proprietary black-boxes?} Today, an independent auditor would need to use arduous reverse engineering techniques to validate devices’ data collection behavior~\cite{Teardown30:online}. 
One alternative is for the device to do everything locally without sending any data~\cite{amini2011cache}, though given that many devices only have basic CPU and storage capabilities, this approach severely limits the functionality they can provide. Doing everything locally also limits many kinds of rudimentary analytics that users might find acceptable, e.g. developers may want to know how many hours the TV is on per week. 
As another alternative, the device can aggregate or denature the data itself before sending it out~\cite{wang2017scalable,yu2018pinto}, but again there is no easy way to verify this behavior. Yet another alternative is to leverage new mechanisms for trusted cloud computing ~\cite{gilad2016cryptonets,konevcny2016federated,lee2019occlumency}. However, these approaches are challenging for users and auditors to understand and verify, and do not necessarily perform data minimization before sensitive data leaves one's control.


\begin{figure*}
	\centering
	\includegraphics[width=\textwidth]{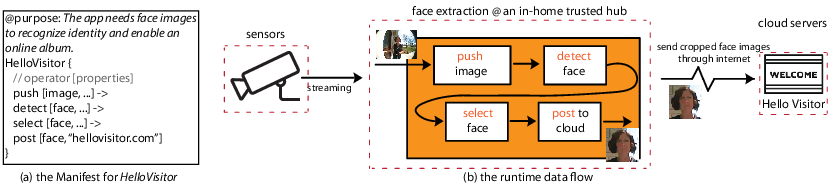}
	\caption{The architecture of the “HelloVisitor” app. A developer creates a manifest making use of four chainable operators (i.e., \textit{push}, \textit{detect}, \textit{select}, and \textit{post}) and adds a text string to specify the purpose. 
	Once deployed, a camera streams the video to the hub, which pre-processes the video so that only portions of images with detected faces are sent to cloud servers. Since the semantics of each operator is known, it is easy to analyze the privacy-related behavior of apps (e.g. “this app only sends images of faces to HelloVisitor.com”) and modify it if desired.}
	\label{fig:sensorhubcloud}
	\vspace*{-0.1in}
\end{figure*}

This paper introduces \name, a new privacy-sensitive architecture for developers to build smart home apps.
\name has three key ideas.
First, app developers must declare all intended data collection behaviors in a text-based manifest (see Fig.~\ref{fig:sensorhubcloud}a), including under what conditions data will be sent outside of the home to cloud services, where that data is being sent to, and the granularity of the data itself. 
Second, to specify these behaviors, developers choose from a small and \textbf{fixed} set of operators with well-defined semantics, authoring a stream-oriented pipeline similar to Unix pipes. This pipeline pre-processes raw data from IoT devices in the home (e.g. sensor data or usage history) into the granularity needed by the cloud service. 
Third, an in-home trusted \name hub mediates between all devices in the home and the outside Internet. This hub enforces the declared behaviors in the manifests, and also locally runs all of the operators specified in these manifests to transform raw data before it is relayed to any cloud services. Combined, these ideas make it so that developers can reduce data collection by running pre-processing tasks on the in-home trusted hub, and users and third-party auditors can inspect data behaviors by analyzing these manifests as well as any actual data flows. Our approach also facilitates a number of privacy features that can be supported by the hub itself, such as adding additional conditions or transformations before data flows out, or transforming parts of the manifest into natural language statements to make it easier for lay people to understand what data will be sent out, when, and to where.

For example, the ``HelloVisitor'' app (Fig.~\ref{fig:sensorhubcloud}) identifies visitors using faces in images. Today’s video doorbells often send captured raw photos to the cloud when they detect a scene change. By applying a pre-processing pipeline (i.e., face detection and image cropping), HelloVisitor can avoid sending images of people or vehicles simply passing by, thus minimizing data egress to just what the app needs to operate.

\name's architecture is based on an analysis (\S\ref{sec:feasibility}) we conducted of 200+ smart home scenarios drawn from the research literature and design fiction interviews we conducted. This analysis led to two insights. First, many apps do not need raw sensor or log data, but rather a transformed or refined version. 
Second, while it cannot support all smart home scenarios, our approach of using a small and fixed set of operators can support a surprisingly large number of scenarios.

We implemented a \name manifest authoring tool on top of Node-Red, a mature, web-based, visual programming platform~\cite{NodeRED78:online}. We developed the operators as a set of Node-Red compatible building blocks. 
Further, we built the \name runtime, which parses a manifest, retrieves raw sensor or log data from IoT devices, sets up a data pipeline by assembling a chain of operators, and streams the data through this pipeline. Note that \name operators are device-agnostic, and rely on runtime drivers to handle heterogeneous device APIs (similar to HomeOS~\cite{dixon2012operating}).
We deployed the \name runtime using a Raspberry Pi connected to a TPU accelerator.

We conducted detailed experiments to validate the design of \name. To understand the range and limitations of its architecture, we first implemented 68 different manifests to cover over 200 use cases and analyzed the types of pre-processing in these manifests (\S\ref{sec:expressiveness}). 
We then used three example manifests to demonstrate the feasibility of using simple algorithms to reduce privacy risks while having little impact on utility (\S\ref{sec:dataegressreduction}). 
We built five end-to-end \name apps, covering 5 data types (video, image, audio, tabular, and scalar) and used these apps to evaluate system performance. We also evaluated the scalability of our low-cost Rasberry Pi setup ($\approx$\$100), \new{showing it can support more than 25 inference tasks and  100 filtering transformations per second} (\S\ref{sec:eval-system}). 
Furthermore, we conducted a user study to test the usability of Peekaboo APIs with 6 developers (\S\ref{sec:zoomstudy}). Our results show that developers can understand how to use \name quickly and complete programming tasks efficiently and correctly. 
Finally, we demonstrated how \name's architecture, and the hub specifically, can support three kinds of privacy features across all apps (\S\ref{sec:independentfeatures}). Specifically, we show how a static analysis tool can generate natural language descriptions and \new{privacy nutrition labels~\cite{DBLP:journals/corr/abs-2002-04631}} of behaviors based on operator pipelines, how an arbitrary manifest can be extended with time-based scheduling features, and how rate limiting can be easily added to a data flow. 


We make the following contributions in this paper: 

\begin{itemize}[noitemsep,topsep=0pt]
	\item A novel software architecture, \name, that helps developers collect sensitive smart home data in a fine-grained and flexible manner, while also making the process transparent, enforceable, centrally manageable, and extensible.
	\item A study of over 200 unique smart home use cases to design a taxonomy of reusable 
	pre-processing operators that can feasibly be implemented at the \name hub to enforce privacy requirements. 
	\item An end-to-end open-source prototype implementation\footnote{\url{https://github.com/CMUChimpsLab/Peekaboo}} of a \name hub on a Raspberry PI platform with a TPU accelerator.   
	\item A detailed evaluation of \name's expressiveness based on coverage of smart home scenarios, system performance, data minimization, and application-independent privacy features.
\end{itemize}

\vspace{-0.1in}
\section{\name Design Overview}

\name has three main components. 
The first is developer-specified \textbf{manifests} that declare \textbf{all} data pre-processing pipelines. Note that the simplest manifest can just describe getting data from IoT devices and sending it to the cloud if no pre-processing is needed. 
The second is a \textbf{fixed} set of reusable and chainable \textbf{operators} to specify these pipelines. 
The third is an in-home trusted \textbf{hub} that enforces these manifests and mediates access between edge devices and the wider Internet. We present more details about our design rationale and tradeoffs below.

\name's manifest is a natural evolution of Android permissions~\cite{felt2012android,nauman2010apex} and Manufacturer Usage Description (MUD) whitelists~\cite{WhatisMU82:online} for IoT devices. 
In Android, developers must explicitly declare permissions in an app's manifest so that it can access protected resources (e.g., location or SMS messages). Similarly, MUDs allow IoT device makers to declare a device's intended communication patterns. The rationale is that many IoT devices are expected to communicate with only a few remote servers known \textit{a priori}, and so declaring the device's behaviors \new{allows the network to blacklist unknown traffic requests.}

However, a weakness with Android's permission system is that it is binary all-or-nothing access. For instance, an app developer might need to access SMS messages from just one phone number for two-factor authentication, but Android only offers access to all SMS messages or none. Furthermore, while the developer can \new{display text} in the app or in a privacy policy that the app will only access messages from one source once, there is no easy way for a user to verify this behavior. 
\new{One solution is to offer many fine-grained permissions for every potential use case, but this would lead to an explosion of permissions that would be onerous for developers to program, unwieldy for users to configure, and complex for platform builders to support.}


\name proposes an alternative approach, requiring developers to declare the data collection behavior inside a text-based manifest \new{using operators}, with the hub only allowing declared flows. 
With \name, a user can install a new smart home app by simply downloading a manifest to the hub rather than a binary.
This approach offers more flexibility than permissions, as well as a mechanism for enforcement. It also offers users (and auditors) more transparency about a device's behavior, in terms of what data will flow out, at what granularity, where it will go, and under what conditions.

\noindent{\textbf{Threat model}}: 
\new{We envision that future third-party developers can build and distribute sophisticated ubiquitous computing apps through smart home App Stores, similar to today's app stores for smartphones. However, these smart home app developers, similar to mobile app developers, might deliberately or inadvertently collect more data than is necessary.}
\name's goal is to limit data egress by such developers, while also making it easier for 
users and auditors to verify the intended behaviors and data practices of their IoT apps.

We make the following assumptions: (1) \textit{Trusted Hub: } The \name Hub, \new{developed by platform providers (e.g., Apple, Google),} is trusted and uncompromised. 
(2) \textit{IoT Devices: } We assume that the \name hub can isolate IoT devices \final{and only allow whitelisted outgoing data flows,} similar to the MUD~\cite{WhatisMU82:online}. All actual IoT Devices are required to send data through the \name hub, and do not circumvent the \name hub, either through an independent or covert side-channel. (3) \textit{Operators: } We assume that the operators that we have created are themselves secure and do not have vulnerabilities. The source code of operators will be made open source and allow for verification, audits, and updates as needed. We note that while our threat model assumes that devices within the home are trusted, it does \textit{not} make a similar assumption about cloud services. 

\section{Analysis of Smart Home Scenarios}\label{sec:feasibility}

To inform the design of \name, we collected over 200 smart home use cases and examined the feasibility and trade-offs of doing pre-processing on a hub. 
Overall, our analysis suggests that, while it cannot support all smart home scenarios, pre-processing data locally before it leaves a smart home via a small set of operators can indeed support a wide range of smart home apps.

\vspace*{0.05in}\noindent\textbf{Collecting smart home use cases}. Since smart home applications are not yet as popular as those on smartphones~\cite{kubitza2020developing}, we chose to study use cases beyond today's commercially available products, sourcing applications through three approaches. First, we conducted a literature review of sensor-based scenarios for smart homes (e.g.,~\cite{larson2012spirosmart}). Second, we surveyed the mobile privacy research literature (e.g., ~\cite{10.1145/2660267.2660270,jin2018they,privacystreams}) to understand common data use patterns by mobile app developers since many of them apply to IoT scenarios. Third, we conducted design fiction interviews~\cite{dunne2013speculative} with 7 participants (3 female, 4 male) across different age groups (min=21, max=60, avg=32) to broaden our set of use cases (See details in~Appendix \S\ref{sec:designfiction}). In total, we gathered over 200 unique smart home use cases. 

\vspace*{0.05in}\noindent\textbf{Method}. 
We clustered similar use cases, resulting in 37 smart home apps spanning different sensors and locations in a home (Appendix Table~\ref{tab:exampleiotapps}). 
We then analyzed what data these apps needed to operate. Specifically, we enumerated why these apps need to collect data (e.g., sensing, analytics, system diagnostics), analyzed potential requirements and constraints in using this data (e.g., compute load, proprietary algorithms, business models), and concluded with what we felt were reasonable outgoing data granularity for each data collection purpose. For example, suppose that the developer of a video doorbell app wants to build an online album for visitors to the home.
A reasonable data requirement of this app is face images. We enumerated these data requirements for different purposes in Appendix Table~\ref{tab:exampledatalevels1}, organizing them by data type.

\vspace*{0.05in}\noindent\textbf{Results}. We make the following key observations. First, \textbf{most cloud services do not need raw sensor or log data}.
For example, an app that uses images to detect potential water leaks likely only needs the parts of the original image showing the floor. A smart lighting app which needs to infer the brightness of a room from a camera only needs derived brightness values rather than the raw image. Table~\ref{tab:exampledatalevels1} offers a quantitative view of data granularity requirements across various scenarios. Only 4 out of 37 camera usages require raw data, while the rest only need either derived or partial data. Beyond cameras, we found that only 19 out of 61 cloud services require raw data, while 8 were scalar values
(e.g., a binary door status sensor). 

Second, \textbf{most pre-processing functions share similar data-agnostic data actions}, suggesting the feasibility of using a small set of reusable operators to replace these repetitive pre-processing implementations. At first blush, the number of pre-processing functions appears huge, since we need to cover various data types, output data granularity (e.g., face, objects, audio events, abnormal data), and filter/transform operations (e.g., image cropping, audio spectrum extraction). However, by enumerating the 200+ smart home uses, we find that there exists consistent, simple, and data-agnostic semantics behind most of the functions. In Table~\ref{tab:exampledatalevels1}, we categorize all the output data granularity into two classes.
\textit{Partial original data} is a subset of the raw sensor data in the original data representation, such as the parts of an image with faces extracted, audio recording segments containing human speech only, and a column in a table. 
\textit{Derived data} is computed from the raw data, often resulting in a new data representation (e.g., the keypoints of a human pose). We use two verbs, ``select'' and ``extract,'' to summarize all the functions that output ``partial original'' and ``derived data,'' respectively.

Third, \textbf{many of these pre-processing functions are lightweight and non-proprietary}, suggesting that they can be run on a low-end smart home hub. For example, pre-trained machine learning models like face detection, functions like inferring brightness, as well as database-like operations on tabular data (project, select, sum, average) are relatively straightforward, openly available, and fast to compute.

\begin{figure*}
	\centering
	\includegraphics[width=\textwidth]{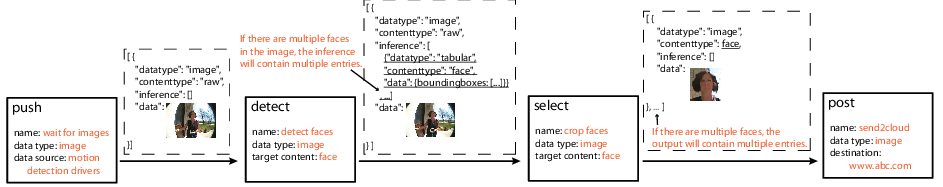}
	\caption{\new{The hub program for ``HelloVisitor'' (Fig.~\ref{fig:sensorhubcloud}).  
	Operators come with a few configurable parameters and have clearly defined ways to interact with a uniform data model (dashed boxes). For example, the \textit{detect} operator only modifies the ``inference'' field. The \textit{select} operator modifies the ``data'' field based on the ``inference'' field. 
    We highlight the affected data fields with underscores. Operators are open source to help with verification of their behaviors.}}
	\label{fig:hubprogramhellovisitor}
	\vspace*{-0.15in}
\end{figure*}

%
\section{Programming Manifests using Operators}\label{sec:peekabooapi}

We discuss the design of \name's manifest and operators, plus the rationale and tradeoffs behind our design. 
The manifest and operators need to be expressive enough to support a wide range of scenarios, easy to author for developers, easy to comprehend for auditors, and beneficial for privacy protection.

\name uses a Pipe-and-Filter software architecture~\cite{garlan1993introduction}, modeling a hub program as a set of connections between a set of stateless operators with known semantics. 
Figure~\ref{fig:hubprogramhellovisitor} presents a data pre-processing pipeline from the \textit{HelloVisitor} app, which can be abstracted into a directed acyclic graph (DAG) of operators. The first operator, a \textit{push} operator named ``wait for images'', specifies the raw image retrieval behavior (e.g., pull v.s. push, frequency, resolution) and gets raw image from the \name runtime when there is a motion event. The second operator, \textit{detect}, annotates the bounding boxes of faces inside the raw image. Next is ``crop face'', a \textit{select} operator that crops the image to output a set of face images based on annotated bounding boxes. Finally, ``send2cloud'' is a \textit{post} operator with outgoing network access, which posts cropped faces to a remote cloud service.

\begin{figure}[h]
	\centering
	\includegraphics[height=3.5cm]{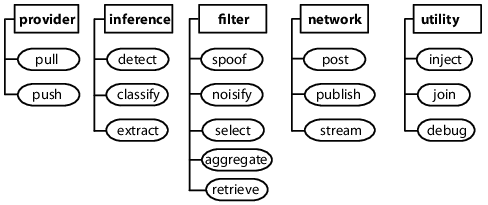}
	\caption{The taxonomy of \name operators. 
	Each operator corresponds to a "verb" statement relative to the operator itself. For example, the \textit{pull} operator pulls data from the hub runtime.
	}
	\label{fig:actiontaxonomy}
	\vspace{-0.15in}
\end{figure}

\subsection{Abstracting Reusable Operators}\label{sec:operators} 

At a high level, pre-processing pipelines do three things: collect raw data from edge devices, transform that data into the targeted granularity, and send the processed data to external servers. Although the exact desired data actions vary across use cases, the high-level data pre-processing semantics are surprisingly similar across data types. For example, a \textit{noisify} data action denatures the original data slightly by imposing some noise, without changing the original data representation format, such as blurring an image, changing pitch/tempo of audio, or distorting numerical values by a small amount. 

We used best practices in API design (e.g.~\cite{bloch2006design}) to guide the design of these operators. We started with a few use cases, programmed them using our initial API, and iterated on the API as we expanded the supported use cases. Based on the data transform behaviors, we created sixteen operators grouped into five categories (Fig.~\ref{fig:actiontaxonomy}): \texttt{provider}, \texttt{inference}, \texttt{filter}, \texttt{network}, and \texttt{utility}. Developers can specify the behavior of an operator by configuring its associated properties.



\begin{figure*}[h]
	\centering
	\includegraphics[width=\textwidth]{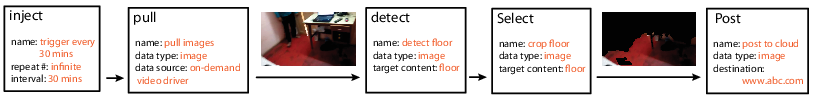}
	\caption{A water leak detection app pulls an image from the camera every 30 minutes, detects the floor area using image segmentation algorithms, and sends the image containing only the floor to the server. The app protects users' privacy by only sending the floor pixels to the cloud.}
	\label{fig:waterleakfloor}
	\vspace*{-0.15in}
\end{figure*}

\vspace*{0.05in}\noindent\textbf{Inferring and filtering target content}. 
In contrast to the binary all-or-nothing data access control, \name aims to enable a new fine-grained semantic-based data access control, which requires developers to declare when the app collects data  (e.g., when a baby is crying) and what data content would be collected (e.g., face images, speech audios).
To achieve this, we introduce two sets of operators: \texttt{inference} operators that can annotate data contents (e.g., the bounding boxes of faces, the key points of a body pose), and \texttt{filter} operators that filter based on the annotations. For example, \textit{HelloVisitor} first uses a \textit{detect} operator to annotate the bounding boxes of faces and then uses a \textit{select} operator to crop the image based on annotated bounding boxes. 

We summarize inference tasks into three primitives: \textit{detecting} instances of objects, \textit{classifying} the dominant content category, and \textit{extracting} derived data (e.g., audio frequency spectrum). 
We also examined common privacy countermeasures~\cite{agarwal2013protectmyprivacy,duckham2006location,6547120,krumm2007inference} to determine five types of \textit{filter} operators: \textit{spoof} for replacing the payload with an artificial replacement, \textit{noisify} for injecting a configurable random noise, \textit{select} for keeping partial raw data, \textit{aggregate} for summarizing statistics, and \textit{retrieve} for overwriting the payload data with derived data.

The abstractions of operators are somewhat analogous to abstract classes in object-oriented programming, and developers can specify the exact data transformation they want via configuring the properties of each operator. 
The runtime then maps the manifest specification to the concrete subclass implementations based on the properties. 
For example, the detect operators in Figs.~\ref{fig:hubprogramhellovisitor} and~\ref{fig:waterleakfloor} have different target content properties, so they are mapped to two different data transformations. Further, there may be multiple face detection operator implementations, and developers can let the runtime determine which one to use or specify one explicitly. 
We enumerate supported data transformations in Appendix Figure~\ref{tab:datatransformations}.

The key property for the inference and filter operators is \textit{target content} (e.g., face in Fig.~\ref{fig:hubprogramhellovisitor} and floor in Fig.~\ref{fig:waterleakfloor}). 
The target content is a type of semantic annotation supported by integrated inference algorithms, which can then be filtered accordingly using filter operators. 
\name currently provides several built-in state-of-the-art inference algorithms using pre-trained machine learning models, which support over 90 visual categories~\cite{chen2017deeplab,lin2014microsoft} (e.g., face, person, floor, table), 632 audio classes~\cite{audioontoloy} (e.g., baby crying), and many other individual categories (e.g., body pose~\cite{Poseesti0:online}, heart rate~\cite{balakrishnan2013detecting}, audio frequency spectrum, brightness). Developers can choose among these options using a dropdown menu in the authoring interface (Fig.~\ref{fig:configinterface}).

Note that a \name runtime only supports a fixed set of operators and property options enabled by the pre-loaded implementations. We do not allow operators to be dynamically loaded because we would not know the semantics of that operator, and it may have undesired behaviors.



\vspace*{0.05in}\noindent\textbf{Collecting raw and relaying processed data}.
To install a \name app, users need to bind the manifest to compatible devices, similar to how users install a SmartThings App~\cite{Onesimpl53:online} today. Developers have to specify required device drivers in \texttt{provider} operators (i.e., \textit{push} and \textit{pull}), so the hub runtime can determine compatible devices.

Here, \textit{push} and \textit{pull} represent two styles of data access: passively waiting for pushed data from drivers and actively pulling from drivers.
For example, \textit{HelloVisitor} (Fig.~\ref{fig:hubprogramhellovisitor}) starts with a \textit{push} operator. 
When there is a significant change in the visual scene, the \name runtime sends an image to the \textit{push} operator to trigger subsequent operators. In contrast, Fig~\ref{fig:waterleakfloor} shows the hub program of a \textit{water leak detection} app, which pulls an image from the camera every 30 minutes.

Finally, network operators are the only group of operators that can send data outside. \name currently has three operators: \textit{post} for HTTP/S post, \textit{publish} for MQTT Pub/Sub, and \textit{stream} for RTSP video streaming. 
Developers can configure the \texttt{provider} and \texttt{network}  operators to enable SSL connections between the IoT devices and the hub, and between the hub and remote servers, similar to the Network security configuration in Android~\cite{Networks78:online}. 



\subsection{Chaining Operators Together}
\label{sec:grammar}

In this section, we present more details on how operators are connected together. 


\vspace*{0.05in}\noindent\textbf{Data model}.\label{sec:detamodels}
\name uses a uniform data structure between operators
(see Figures~\ref{fig:hubprogramhellovisitor} and \ref{fig:waterleakfloor}).
The basic data structure (i.e., a \name data item) is a map, and the message transmitted between operators is an array of such items. An example map is shown below.
\begin{align*}
	\begin{split}\label{datastructure}
 \{``datatype"&: ``video\, | \, image \, | \, audio \, | \, tabular \, | \, scalar", \\ ``contenttype"&: ``raw \, | \, face \, | \, person \, | \, dog \, | \, speech \, | \, ...  ", \\
 ``inference"&: [  \{...\}  \, | \,   \, \{...\} \, | \, ... ],\\``data"&: \{raw \, sensor \, data\},\\``process"&: \{device \, information,  operation \,  history\}  \} \\
 \end{split}
\end{align*}
A \name data item only stores one unit of the data (e.g., an image, audio file, tabular row, or scalar value), while the inference field contains a list of annotations. Suppose the input of  \textit{HelloVisitor} is an image with multiple faces. Here, \textit{detect face} will write multiple face annotations to the inference field, and each face annotation is a tabular \name data item. \textit{crop face} will then generate a list of \name data items, where each corresponds to one face image (Fig.~\ref{fig:hubprogramhellovisitor}).


\vspace*{0.05in}\noindent\textbf{Composing pipelines}.  
To build a data pre-processing pipeline, developers connect operators' outputs to others' inputs and assemble them into a directed acyclic graph. 
All operators, except \textit{join} (described more below), take only input from one prior operator. This design avoids synchronization issues since the execution of operators is asynchronous. 

In contrast, developers can connect an operator’s output to multiple operators’ inputs. The preceding operator creates a copy of the output message for each connection to avoid interference and potential multithreading problems.




\begin{figure*}
	\centering
	\includegraphics[width=0.95\textwidth]{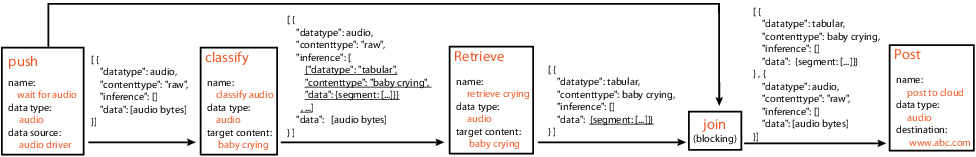}
	\caption{A baby monitoring app only sends audio containing a crying sound to the server. \new{The \textit{retrieve} operator replaces the ``data'' field with the ``inference`` field.
    The \textit{join} operator merges the outputs from \textit{wait for audio} and \textit{retrieve crying}, passing its output onward when both input streams arrive. Finally, the \textit{post} operator  only sends the audio data item outside, since its data type parameter is ``audio''.}}
	\label{fig:appbabycrying}
	\vspace*{-0.2in}
\end{figure*}

\vspace*{0.05in}\noindent\textbf{Supporting more complex logic}. 
Figs.~\ref{fig:hubprogramhellovisitor} and \ref{fig:waterleakfloor} show two simple pipelines. However, these linear pipelines are insufficient for many applications. Imagine a baby monitoring app that only sends input audio to an external server when the hub program detects a baby crying (Fig.~\ref{fig:appbabycrying}). With a linear pipeline this is infeasible since when we use a \textit{retrieve} operator to check if the audio contains baby crying, the subsequent operators no longer have access to the original data. More fundamentally, this is a common constraint of many dataflow programming models, where the data itself controls the program's flow. 

We address this limitation by introducing the \textit{join} operator to support AND, OR, NOT logic. \textit{Join} is the only operator that can take input from multiple operators and fuse asynchronous data flows into one pipeline. A key property for \textit{join} is whether it is blocking or non-blocking. A non-blocking \textit{join} forwards incoming data whenever it becomes available, equivalent to an \textbf{OR} logical operator. In contrast, a blocking \textit{join} only merges and forwards incoming messages if they meet specified criteria (e.g., all the incoming messages arrive within a small time window). Note that incoming messages can be from the same prior operator. For example, a blocking \textit{join} can block the data flow until there are multiple ``baby crying'' events, to confirm that a certain accuracy level is met. 


\vspace{-0.2cm}
\subsection{Error Handling \& Debugging Support} 
It is possible for the flow between operators to be \name data items with different data types due to flow fusion. Without careful design, connecting arbitrary operators together may result in unpredictable errors. An essential property of our API design is that each operator is associated with a target data type. An operator only processes corresponding data items selectively, so the operators will not try to detect faces in a scalar value object. 

Another important design issue is that \textit{inference} and \textit{filter} operators handle unmatched data types differently. Inference operators will leave unmatched items untouched, while filter operators will filter them out. This design strengthens \name's annotation capability since developers now can apply multiple inference operators to the same data item. Meanwhile, it enhances privacy as well: data can flow to the next operator only if the developer matches the data type explicitly. 

To help with debugging, developers can append a \textit{debug} operator after any operator, which will print output data to a console window. 
We also incorporated example data (e.g., test videos, photos, audios, and tabular data) as options of the \textit{pull} operator and designed the \textit{inject} operator to support both manual and interval triggers, so developers can quickly test pipelines within the editor without actual hardware.

\section{Illustrating Pre-processing using Examples}\label{sec:casestudies}


\begin{figure*}[t]
	\centering
	\includegraphics[width=\textwidth]{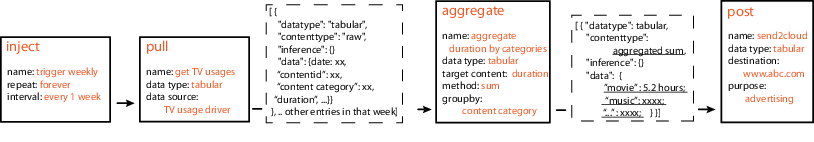}
	\caption{A SmartTV usage summary app queries users' watching patterns and uses the data to generate a weekly summary for video consumption across different channels and categories. 
	\new{The \textit{aggregate} operator works similarly to SQL aggregation and replaces the value in the ``data'' field.}} 
	\label{fig:smarttvaggregation}
	\vspace*{-0.15in}
\end{figure*}

\begin{figure*}[t]
	\centering
	\includegraphics[width=\textwidth]{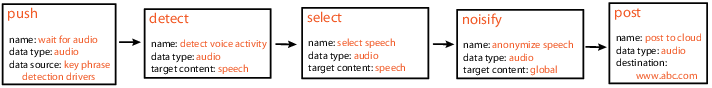}
	\caption{The {incognito} voice assistant only collects short audio segments containing speech and hides the speaker's identity by changing the pitch. This hub program can protect users' identity without breaking the speech recognition functionality.
	}
	\label{fig:app-voiceassitant}
	\vspace*{-0.15in}
\end{figure*}

\begin{figure*}[t]
	\centering
	\includegraphics[width=\textwidth]{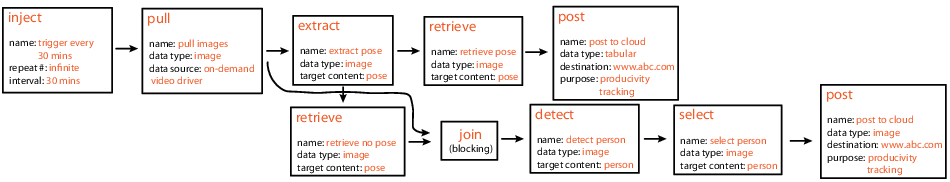}
	\caption{The productivity tracking app pulls an image from the camera every 30 minutes, analyzes the pose of the person, and uploads the pose to the server for further analysis. Pose information is usually unavailable when the person sits at a desk. The app then only sends the person pixels to the cloud with more sophisticated algorithms.}
	\label{fig:app-productivity}
	\vspace*{-0.1in}
\end{figure*}

We present three example manifests to illustrate the use of \name's APIs. 

\vspace*{0.05in}\noindent\textbf{Smart TV logs}. A smart TV developer is interested in collecting users' viewing history for advertising purposes. This smart TV stores viewing history in a simple tabular form.
Here, we assume developers want to do better with respect to privacy, perhaps for legal compliance, market competition, or because hubs like \name are widely adopted in the future and companies want to assure customers that they are only collecting minimal data. Fig.~\ref{fig:smarttvaggregation} shows an example pipeline we built to show how a developer can compute an aggregate view of video consumption on the hub, thus sending out a less sensitive summary. This example also shows \name's support for database-like queries (e.g., SELECT, AGGREGATE, JOIN, WHERE), which also work in other tasks (e.g., counting people in an image).


\vspace*{0.05in}\noindent\textbf{Incognito voice assistant}. 
\final{Fig.~\ref{fig:app-voiceassitant} presents a manifest of a smart voice assistant that we developed, which can offer a speech anonymization feature that protects users from exposing undesired voice fingerprints. 
In contrast to Google's ``Guest Mode''~\cite{GuestMod65:online}, 
this manifest can \textbf{assure} users that their voice fingerprint identities are protected.}

Beyond database-like queries, \name introduces two important extensions, content-based selection and explicit noise injection, to accommodate the smart home context and privacy-preserving goals. This example illustrates how the combination of inference and filter operators can filter out non-speech audio segments.
Further, this example uses a \textit{noisify} operator to hide speaker identity, which changes the tempos and pitch of the captured audio with a configurable random variation (e.g., 5\%).


\vspace*{0.05in}\noindent\textbf{Productivity tracking}.\label{sec:productivity} PC Applications like RescueTime~\cite{RescueTi20:online} help users be more productive by helping them understand how they spend their time. As working from home becomes increasingly common, we envision a smart home version 
that track a person's productivity beyond PCs. Implementing such an app in a conventional cloud-based architecture can be worrisome due to privacy concerns similar to those in the voice assistant app. 
Fig.~\ref{fig:app-productivity} presents an example pre-processing pipeline of a productivity tracking app using a camera, which transmits extracted poses to the cloud. 

This example shows use of \name's flexible conditional flow control. 
\name operators support an intrinsic condition flow control similar to Unix Pipes. 
The runtime executes an operator only if its input is ready, and the filter operators only forward non-empty processed data to subsequent operators. For example, if the retrieve pose operator (Fig.~\ref{fig:app-productivity}) cannot find any extracted poses in its input, the flow stops propagating. This simple design makes it easy for \name APIs to support IFTTT-like trigger-action programs natively.

This example also shows how the output of a single operator can be forwarded to multiple operators, similar in spirit to Unix's \textit{tee} command. The blocking and non-blocking \textit{join} operators allow \name to accommodate distributed asynchronous sensors, supporting smart home apps with complex logic across distributed devices within the same home. 

\section{Implementing the Hub Runtime}
\label{sec:implementation}


\begin{figure*}
	\centering
	\begin{minipage}{.62\linewidth}
		\includegraphics[width=\columnwidth]{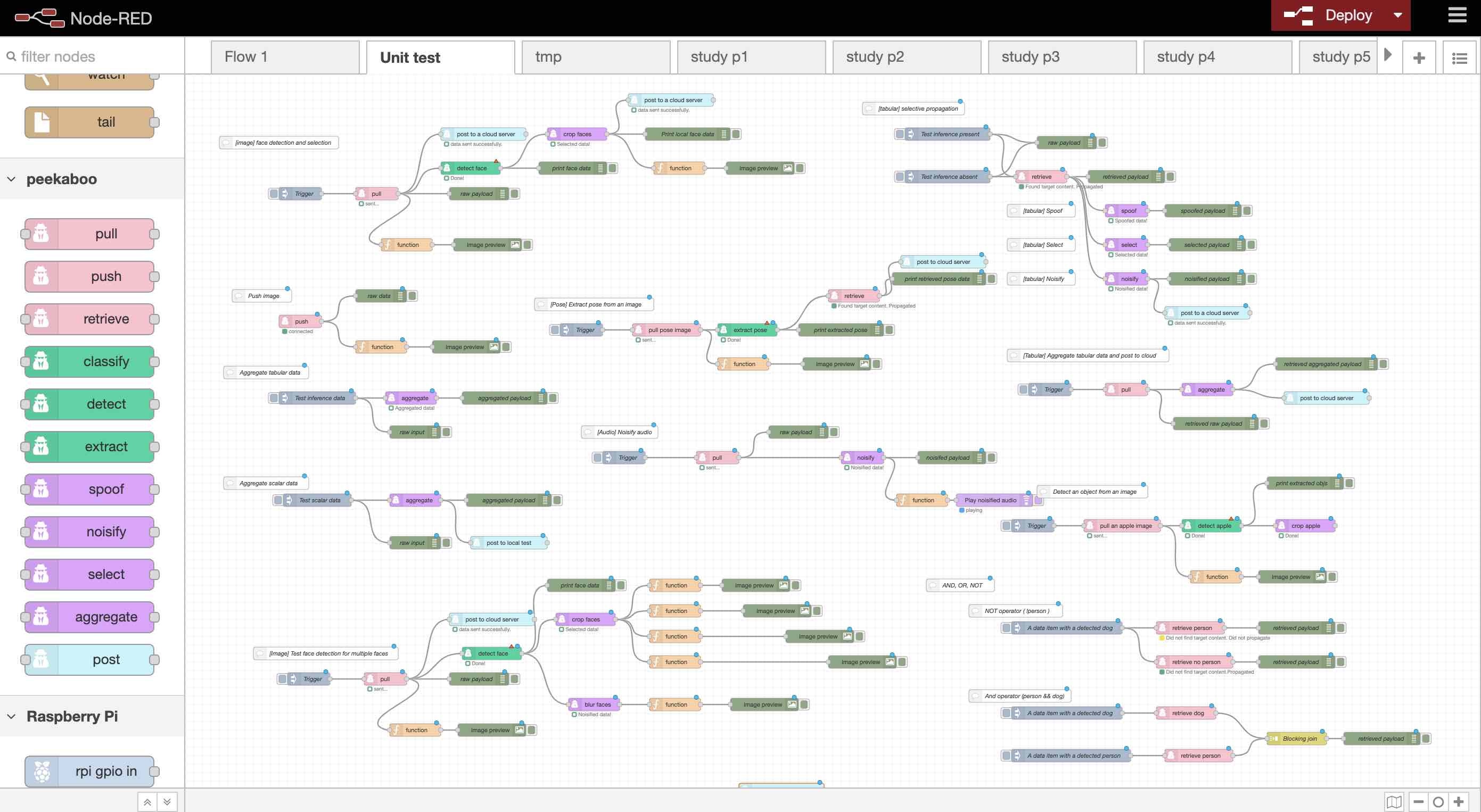}
		\caption{The editor interface and the unit test tab presented to the developers. Developers can inspect the unit tests to understand operators' behavior and the connection grammar. }
		\label{fig:webinterface}
	\end{minipage}
	\hfill
	\begin{minipage}{.36\linewidth}
	\includegraphics[width=\columnwidth]{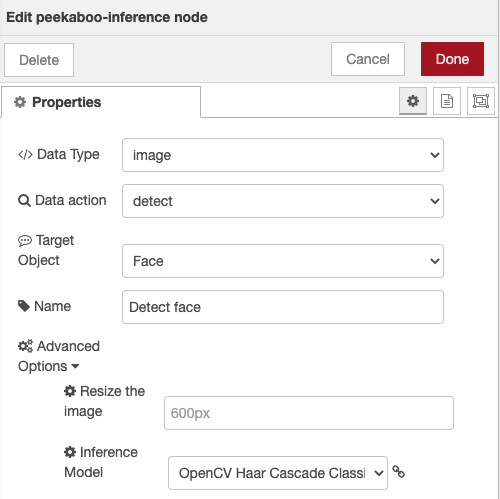}
	\caption{The configuration interface of an inference operator.}
	\label{fig:configinterface}
	\end{minipage}
	\vspace{-0.2in}
\end{figure*}

We implemented \name by leveraging Node-Red~\cite{NodeRED78:online}, a visual programming platform developed by IBM to wire together devices using a library of customized blocks. The notion of ``blocks'' and directional ``connections'' in the visual programming language are well suited to \name's chainable operators. We implemented \name operators as a set of Node-Red compatible building blocks, so developers can use a Node-Red web-based flow editor UI as the programming environment and leverage its built-in debugging utilities (e.g., displaying images, playing an audio, printing output data). 
\final{The source code is available at \url{https://github.com/CMUChimpsLab/Peekaboo}.}

\vspace*{0.05in}\noindent \textbf{Operators: } Each \name operator is written in JavaScript and executed by a Node-red runtime once deployed. One challenge in \name is the relative lack of support for machine learning algorithms in JavaScript. Our experiments with multiple JS implementations (e.g., opencv4nodejs) showed them to be slow and inaccurate. Instead, we design the \name runtime using a microservice architecture, where several ML inference algorithms are hosted in containerized services running on the \name hub, and only \name inference operators can communicate with these services locally through web sockets.

\vspace*{0.05in}
\noindent \textbf{Drivers to obtain sensor data: } 
Each \name hub program gets data from hub runtime drivers rather than querying hardware directly. Developers may implement customized runtime drivers in arbitrary code, similar to HomeOS~\cite{dixon2012operating}. 
These drivers retrieve raw sensor data (video, image, audio, tabular, or scalar) from edge devices, format it 
(e.g., b64string for images, bytes for video/audio, JSON for tabular data), and publish the data in a local socket. The hub program can either pull the latest data (e.g., retrieving a real-time photo) from the driver or register a push subscription on the driver (e.g., receiving a photo if there is a significant scene change). Note that these runtime drivers handle device heterogeneity and specifics of getting data from them, while the \name operators are device-agnostic data stream operators. 


\vspace*{0.05in}\noindent \textbf{Runtime:} We deployed the \name runtime (i.e., Node-Red, drivers, containerized inference services) on a Raspberry Pi 4B (4GB), along with a Google Coral TPU and a Zigbee adaptor, with a total cost of $\approx$ \$100 USD.
Most of our deep-learning-based inference tasks use pre-trained MobileNet~\cite{sandler2018mobilenetv2} models and are outsourced to the TPU. 

\vspace*{0.05in}\noindent \textbf{End-to-end applications:} 
We developed drivers for four \name IoT devices, a customized smart camera and smart speaker using Google AIY Kits~\cite{AIYProje53:online}, a simulated RESTFul API that generates tabular smart TV logs, and an Aqara Zigbee humidity sensor, covering the five data types in Table~\ref{tab:exampledatalevels1} (video, image, audio, tabular, and scalar).

We built five end-to-end applications using these devices: video based heart rate measurement~\cite{balakrishnan2013detecting}, HelloVisitor (Fig.~\ref{fig:sensorhubcloud}), incognito Speech Assistant (Fig.~\ref{fig:app-voiceassitant}), Smart TV logs collector (Fig.~\ref{fig:smarttvaggregation}), and humidity-based irrigation reminder. We built customized cloud services and web UIs for all five apps.

\begin{table*}[h]
	\renewcommand{\arraystretch}{1.2}
	\begin{center}
		\captionof{table}{We implemented 68 unique manifests for over 200 smart home use cases, analyzed the types of pre-processing in these manifests, and if the data needs for that use case could not be supported, we examined why. 
		}
		\small 
		\begin{tabular}{ |p{30mm} | p{30mm} | p{110mm} | }
			\hline	
			Pre-processing & \#Scenarios supported & Why \name could not support some of the scenarios \\ \hline  \hline
			Content selection &  64 / 68 & e.g., hard to select content of interest, need to be used for online albums  \\ \hline 
			Conditional filtering & 57 / 68 & e.g., high-stake tasks, proprietary implementations, insufficient computing resources \\ \hline 
			Explicit noise injection & 51 / 68  & e.g., high-stake tasks, injected noise may break the intended tasks \\ \hline \hline 
			Always need raw data & 3 / 68  &  the intersection of the three categories listed above \\
			\hline 
		\end{tabular}
		\label{tab:notsupporttable}
	\end{center}
	\vspace{-0.2in}
\end{table*}


\section{Evaluation}\label{sec:evaluation}

This section presents detailed experimental evaluations of the feasibility and benefits of \name. We first created 68 different manifests for all the 200+ smart home use cases (Section \ref{sec:feasibility}). Note, these manifests only work with bare-bone server implementations as we do not have proprietary backend implementations from manufacturers. We then analyzed types of pre-processing opportunities that can be enabled by these manifests (\S\ref{sec:expressiveness}). 
We also investigated the feasibility of balancing the tradeoff between privacy protection and utility for each type of pre-processing (\S\ref{sec:dataegressreduction}). 
We later built five end-to-end real-world \name apps, covering five data types and used these apps to evaluate our system performance (i.e., latency and throughput) (\S\ref{sec:eval-system}).
Finally, we demonstrate several hub privacy features that both developers and users essentially get for free, which work across apps and devices (\S\ref{sec:independentfeatures}). 

\subsection{Evaluating \name's coverage of smart home scenarios}\label{sec:expressiveness} 

To validate the feasibility of \name, we authored manifests to cover the use cases described in \S\ref{sec:feasibility}.

\vspace*{0.05in}\noindent\textbf{Method}. A smart home use case can be realized through different types of sensors. For example, a dedicated occupancy sensor or a camera can enable an occupancy-based application, although required data collection behaviors can differ. 
Two authors collaboratively implemented one manifest for each usage-sensor pair to explore different pre-processing opportunities. We found that many manifests could be reused for multiple scenarios, and so this process resulted in 68 manifests.

\vspace*{0.05in}\noindent\textbf{Results}. Of the 68 manifests, we only identified three that always need raw data: a smart cooking device that uses a camera to analyze ingredients, a smart toilet that analyzes poop for diseases, and a microphone that performs spirometry~\cite{larson2012spirosmart}. 
For these kinds of apps, developers can directly connect a \texttt{provider} operator to a \texttt{network} operator to send out the raw data. While \name does not reduce data egress in these cases, it still provides transparency of data collection (e.g., what data has been sent to the service provider, and how often) and a binary on-off control.
 
For the other 65 manifests, \name APIs can enable at least one of the following types of pre-processing: content selection (e.g., cropping faces), conditional filtering (e.g., only sending data if a person appears in the view), and explicit noise injection (e.g., changing the pitch of an audio recording).
Table~\ref{tab:notsupporttable} enumerates the breakdown of supported pre-processing across all scenarios, as well as the unsupported reasons.


Content selection is the most common pre-processing (64 of 68 scenarios). Examples include cropping faces from images, extracting audio frequency spectrum, and aggregating numerical values. 
We could not apply content selection to 4 apps for two reasons. First, the algorithms to select the content of interest cannot run on a \name hub. For example, the algorithm to recognize food ingredient is proprietary and may require significant computational resources and frequent model updates. Second, developers need raw data to fulfill the data collection purpose. For example, automatic photography (e.g., Google Clips~\cite{GoogleAI96:online}) needs to collect and store the original photo.

Conditional filtering is also common (57/68). Since many smart home apps are event-driven, adding a local event filter can significantly reduce data egress. For example, a manifest that filters images based on the presence of faces can reduce the number of outgoing images from a basic motion-activated camera. However, one constraint of \name's architecture is that the open-source hub algorithms may not be as accurate as their cloud counterparts. As a result, we cannot insert filters for high-stake tasks, e.g., elderly fall detection. Another constraint is that some conditional filters might be proprietary with no current open source equivalent (e.g., water leakage detection). Finally, some conditional filters may require extensive computational power and storage, which cannot run on a local hub (e.g., wanted criminal search using smart doorbells).

While it is possible to inject explicit noise into the data pre-processing pipelines, there is one crucial privacy-utility trade-off: the injected noise may break the intended functionality and reduce the service quality. So we cannot explicitly inject noise to apps for high-stake tasks. Besides, many scenarios collect only coarse data (e.g., binary occupancy) or need raw data, where explicit noise injection is not applicable.



\subsection{Privacy-utility trade-offs}\label{sec:dataegressreduction}

While some data transformations are unlikely to affect service quality (e.g., many kinds of tabular data aggregation), others might have negative impacts. A main concern for content selection and conditional filtering is that the open-source inference models on the hub may be less effective than large proprietary machine learning models on the cloud. Furthermore, for explicit noise injection, that noise might break intended functionalities.
While the actual trade-offs depend heavily on the applications, we used three example apps to demonstrate the \new{\textbf{feasibility}} of using simple algorithms to reduce privacy risks while having little impact on utility.

\vspace*{0.05in}\noindent\textbf{Method}. 
We chose three pre-processing tasks to evaluate: a face-only video doorbell (HelloVisitor, Fig.~\ref{fig:sensorhubcloud}, content-selection), a person-activated camera (Fig.~\ref{fig:app-productivity}, conditional filtering), and an incognito voice assistant (Fig.~\ref{fig:app-voiceassitant}, noise injection). 
We chose these tasks because they represent different types of pre-processing, the availability of labeled ground truth data, and the availability of publicly accessible cloud-based baselines. 

The face-only video doorbell can improve privacy by only sending images of faces, although \name's operator might miss some faces that a more capable cloud algorithm could detect. Similarly, the person-activated camera reduces unnecessary outgoing images, but it may miss images that potentially contain a person. For these two apps, we quantified the privacy benefits using the percentage of data egress reduction in bytes and the impact on utility using F1 scores. Finally, our incognito voice assistant protects speakers' identity by changing the tempo and pitch of the captured audio with small random variations using the \textit{noisify} operator (<10\%). We quantify the privacy benefit by measuring speaker recognition error and the utility impact using speech recognition error. We measured accuracy using the Levenshtein function~\cite{Levensht34:online} to compare the recognized speech text with the ground truth text.

We ran the experiments with multiple benchmark data sets. First, we used the ChokePoint dataset~\cite{wong2011patch}, a person identification dataset under real-world surveillance conditions, to test the face-only video doorbell. Next, we selected 457 ``home office'' videos from an in-home activity dataset~\cite{sigurdsson2016hollywood} to evaluate the person-activated cameras. Finally, we used speech recordings from the CMU PDA database~\cite{PDAspeec59:online} to test the incognito voice assistant \new{(6 unique speakers, 112 unique audio files). For each speaker, we used half of the audio files (7-15 files) for speaker enrollment and the other half for speaker recognition and speech recognition.} We used state-of-the-art cloud-based solutions from Microsoft Cognitive Services (i.e., face detection, person detection, speaker recognition, speech recognition) as the baseline~\cite{Cognitiv48:online}.

\begin{table}[h]
	\renewcommand{\arraystretch}{1.2}
	\begin{center}
		\captionof{table}{Simple pre-processing algorithms provide significant improvement in the amount of potentially sensitive data sent, while maintaining utility.}
		\small 
		\begin{tabular}{ |p{35mm} | p{20mm} | p{20mm} | }
			\hline	
			 & Privacy Metric &  Utility Metric \\
			 \hline \hline
			 & Outgoing data & F1 Score \\ \hline
			 \name face doorbell & \textbf{6.0}\% & 94.3\% \\ 
			 Baseline doorbell & 100\% & \textbf{95.3}\% \\\hline \hline
			 \name person camera & \textbf{4.3}\%  & 93.2\% \\
			 Baseline camera & 100\% & \textbf{96.2}\% \\ \hline \hline
			 & Speaker Recog Accuracy & \new{Speech Word Error Rate} \\ \hline
			 \name  voice assistant & \textbf{27.7}\% & \new{11.88\%} \\
			 Baseline voice assistant & 100\% & \new{\textbf{9.27}\%} \\
			 \hline \hline
		\end{tabular}
		\label{tab:privacyutilitytradeoff}
	\end{center}
	\vspace{-0.2in}
\end{table}


\begin{figure*}[t]
	\centering
	\includegraphics[width=\textwidth]{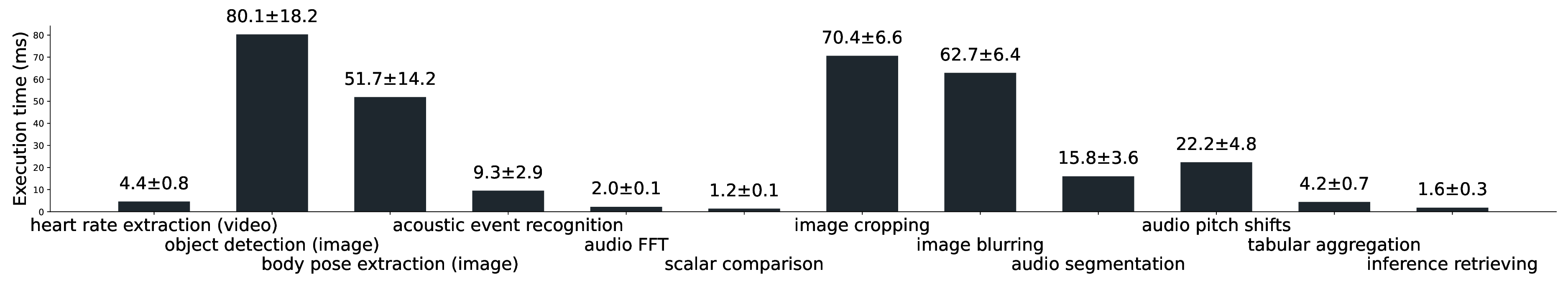}
	\caption{The execution times (in milliseconds) for transforming 1-second videos/audios (normalized), 800x600 images, and a 5-column table with 100 entries. Our low-cost setup can support 25 inference tasks and more than 100 filtering transformations per second, which we believe is sufficient to support many smart home scenarios.}
	\label{fig:executiontime}
	\vspace*{-0.15in}
\end{figure*}


\vspace*{0.05in}\noindent\textbf{Results}. Table~\ref{tab:privacyutilitytradeoff} presents the privacy-utility tradeoffs across our three example apps. The face-only camera reduces data egress from 1 million full-resolution images (12 GB) to 64,000 face images (200 MB), while the F-1 score of the local model is only 1\% lower than the Microsoft Azure API. 
The person-activated app replaces 2868 images (62.8\%) with less privacy-sensitive pose key points and removes unnecessary background pixels from 770 images (16.8\%), resulting in a data reduction of 95.7\% (from 164 MB to 7 MB). Meanwhile, the F-1 score of the local model is only slightly lower than the Microsoft Azure API (93.2\% v.s. 96.2\%). The incognito voice assistant reduces  speaker identification accuracy from 100\% to 27.68\% (lower is better since it provides more anonymity) while only reducing speech recognition accuracy by \new{2.61\%}.
Our results show that \name's pre-processing (e.g., small random pitch shifts, pre-filtering) can significantly reduce data egress with minor adverse impacts on the intended tasks.




\subsection{System Performance}
\label{sec:eval-system}

\name's architecture has two major constraints. The first is that the hub has more limited computing resources than cloud servers. 
Resource constraints on the hub may lead to pre-processing of data taking longer, or limit the  ability to scale to many simultaneous pre-processing tasks in a smart home. Second, the pipe-and-filter architectural style introduces latency due to repetitive parsing and unparsing across filters. We conducted experiments to quantify the \textit{computation load (i.e., throughput)} of data transformations and the \textit{end-to-end latency} of different pipelines. We note, however, that our prototype is not highly optimized and is running on relatively low-end hardware, so this evaluation is intended to provide a rough lower bound on performance.

\vspace*{0.05in}\noindent\textbf{Method}. We deployed the hub runtime to a low-cost setup and used the four sensors as the edge devices, both described in \S\ref{sec:implementation}. We also set up a cloud server on an AWS p2.xlarge instance, in an AWS region close to the authors institution.

We first characterized the computation load of common data transformations identified in~\S\ref{sec:expressiveness}, including both inference and filtering. For inference, we profiled object detection (e.g., face, person, floor), audio event recognition (e.g., baby crying, water dripping), audio frequency spectrum extraction, body pose extraction, video heart rate extraction and scalar value comparison. For filtering, we profiled object-based image cropping, object-based image blurring, audio segmentation, audio pitch random shifts, tabular data aggregation, and inference results retrieving. 

We used 3 videos from Intel's IoT Devkit~\cite{inteliot90:online} (\char`\~2 min) to test video transformations, 3 sound clips from Google AudioSet~\cite{audioontoloy} (\char`\~10 seconds) to test audio transformations, 10 images from the ChokePoint dataset~\cite{wong_cvprw_2011} (800x600) to test image transformations, and a synthetic dataset containing 100 entries to test tabular data transformations. We then measured the average computation time of 1000 repetitions.


Next, we compared the difference in latency when  running pre-processing tasks on the hub vs running them on the cloud. With respect to the former, we measured the pre-processing latency on the hub and the transmission time of sending the processed data to the cloud. With respect to the latter, we measured the transmission time of sending the raw data to the cloud and the pre-processing latency on the cloud. 
We used the dataset mentioned above to test 3 end-to-end apps described in \S\ref{sec:casestudies}. \final{According to prior work~\cite{apthorpe2017smart}, the frequency of network events from typical sensors (e.g., sleep monitors, nest cameras, switches) varies from a few per minute to a few per hour. We stress tested the requests} at an interval of 0.5 seconds and measured the average latency of 1000 repetitions.

\begin{figure}[h]
	\centering
	\includegraphics[width=\linewidth]{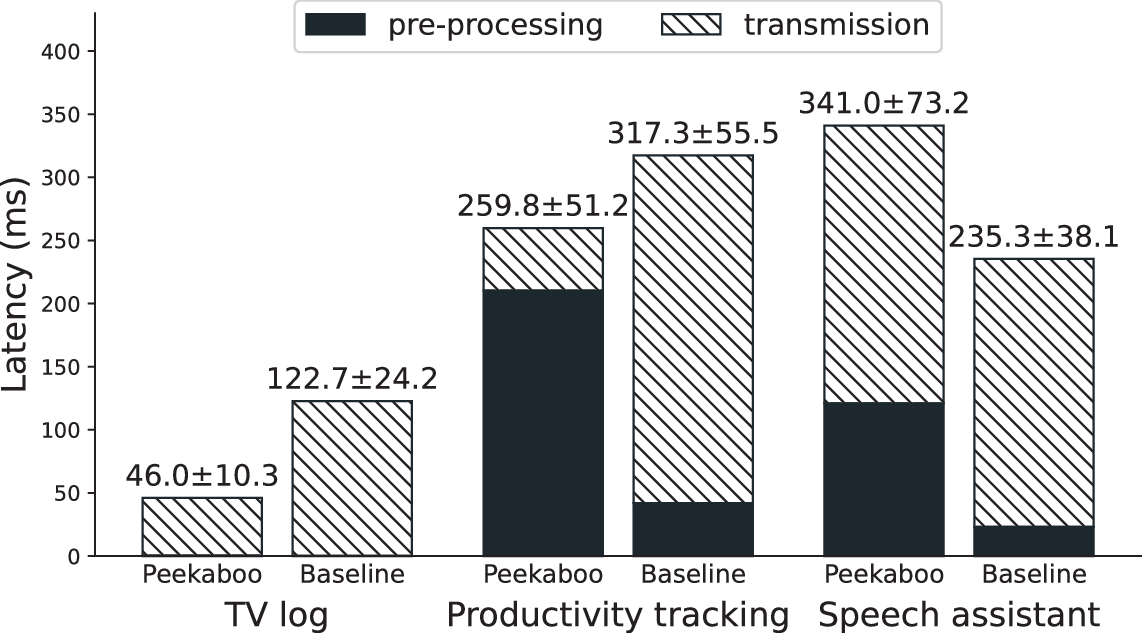}
	\caption{The latencies for the three apps are comparable to the conventional cloud-based approaches. The pre-processing times for TV logs are negligible. Although \name apps spend more time pre-processing data on the hub than on the cloud, they may spend less time on data transmission since pre-processing can reduce outgoing data size.}
	\label{fig:endtoendlatency}
	\vspace{-0.05in}
\end{figure}

\vspace*{0.05in} \noindent\textbf{Results}:
Figure~\ref{fig:executiontime} presents the individual completion times on a Raspberry Pi 4B for common data transformations. Most filtering tasks take 5 ms to 80 ms to complete, while the inference tasks on multimedia data are generally more expensive. Although inference tasks (e.g., object detection) take around 80 ms to complete, they consume little CPU resources since the core computation are outsourced to the TPU device, and the operator runs asynchronously. On average, most of the ML models we use (MobileNet~\cite{sandler2018mobilenetv2}) take around 40 ms on the TPU per inference. So we estimate that our low-cost setup can support 25 inference tasks and more than 100 filtering transformations per second, which we believe is sufficient to simultaneously support many smart home scenarios. 

Figure~\ref{fig:endtoendlatency} presents the average latency for different apps, showing that \name can achieve a latency comparable to conventional cloud-based approaches. The TV log app's hub program spends negligible time on log aggregation but reduces the outgoing data size significantly, thus experiencing lower latency with \name. Similarly, \name's productivity tracking app spends 210.2 ms (std=49.6 ms) to extract the pose, reducing the network transmission time from 275.3 ms (std=50ms) to 40.1 ms (std=11.1ms). The \name speech assistant app takes 106ms more to pre-process and send a 10-second sound clip to the cloud than the conventional approach since the pre-processing in this case does not reduce the outgoing data size. 
In summary, the latency for pre-processing outgoing data on the hub is comparable to the conventional cloud-based approach. Some \name apps have reduced data transmission time due to the reduced size of outgoing data.

\subsection{Developer User Studies}\label{sec:zoomstudy}

\vspace*{0.05in}\noindent\textbf{Method}: 
To evaluate the complexity of creating hub programs using operators, we conducted an IRB-approved study with six developers (3 Male, 3 Female, ages 21-26). Participants were asked to author four manifests using an IDE accessible from their web browsers. 
We configured the inference services and drivers on our hub as well as all cloud services in advance of the study, because we wanted to focus on the usability and understandability of \name's programming model.

All participants had at least three years of programming experience, 3 have developed mobile apps, and 2 have developed apps for IoT scenarios specifically. None had used the Node-Red IDE before. Each study took around 60 minutes, and we paid each participant \$15 upon completing the study. The study was performed remotely over Zoom video conferencing.

During the study, we provided our participants with limited overall assistance -- a brief introduction to \name and the Node-Red interface, some basic documentation on operators (i.e., the taxonomy in Fig.~\ref{fig:actiontaxonomy}), many unit tests for each operator with different properties (Fig.~\ref{fig:webinterface}), and the \textit{HelloVisitor} program as a complete example serving as a warm-up task. 
Participants can inspect and run the unit tests to gain a concrete understanding of the input and output of each operator and the connection grammar.  

We presented each participant with four tasks: the first three were the three case studies (\S\ref{sec:casestudies}), and the fourth was an additional open-ended task of their choice. We described the context and the desired data granularity (i.e., data content and conditions) for each task and asked them to implement the corresponding hub programs. We did not randomize the order, since we found that ordering the tasks based on difficulty can scaffold participants' learning process in our pilot studies.  Finally, we marked the task completed if the hub program could send the right data to a pre-configured server.

\begin{table}[t!]
	\renewcommand{\arraystretch}{0.99}
	\begin{center}

		\small 
			\begin{tabular}{ || p{0.27in} || p{0.07in} p{0.24in} | p{0.07in}  p{0.3in} | p{0.07in}  p{0.3in} | | p{0.07in}  p{0.3in}  || }
			\hline	\hline	
			\multirow{2}{0.27in}{Parti-cipant } & \multicolumn{2}{c|}{Task 1} & \multicolumn{2}{c|}{Task 2} & \multicolumn{2}{c|}{Task 3}  & \multicolumn{2}{c|}{Task 4} \\ \cline{2-9}
			 &Time   & ~~~\# & Time   & ~~~\# & Time   & ~~~\# & Time   & ~~~\#    \\ \hline 
			 \#1 & 5 & 6 (1)  &  6 & 9 (1) & 10  & 17 (4) & 7 &  15 (5) \\ \hline 
			  \#2 & 4  &   6 (2) &  4  &  10 (2) & 12  & 16 (5) & 2 & 4 (0)\\ \hline 
  			  \#3 & 5 & 5 (1)  &  4 & 6 (0) & 9  & 16 (4) & 3 & 8 (0) \\ \hline 
			  \#4 & 2 & 6 (2)  &  4 &  12 (3) & 21 & 9 (4) & 2 & 5 (0) \\ \hline 
 			  \#5 & 3 & 6 (2) &  5 & 9 (1) & 12  & 15 (5) & 2 & 4 (0) \\ \hline 
  			  \#6 & 3 & 4 (1)&  4 & 7 (1) & 7  & 12 (2) & 2 & 5 (0) \\ 
			\hline  \hline
		\end{tabular}
				\captionof{table}{All participants learned the APIs quickly and completed the tasks successfully. The ``Time'' column shows minutes spent on each task. The ``\#'' column contains total numbers of operators plus \textit{debug} operators (in parens) in the final hub program.  After the first three manifests, most participants implemented manifests without using debug operators.}
		\label{tab:taskcompletion}
	\end{center}
	\vspace*{-0.30in}
\end{table}

\vspace*{0.05in}\noindent\textbf{Results}: 
Table~\ref{tab:taskcompletion} shows the quantitative results of our lab study. All participants completed the tasks successfully. The task completion time across participants are similar: 3-5 minutes for task 1, 4-6 minutes for task 2, and 7-21 minutes for task 3, reflecting the difficulties of these tasks. Most participants picked up \name's programming model quickly without much explanation. For example, participants \#4, and \#5 completed Task 1 without using the unit tests. 

We also count the total number of operators, and the number of \textit{debug} operators, in the final hub programs. In the first three tasks, most participants actively used the debug operator to help them understand each operator's behavior. When they reached the fourth task, they were quite confident about using these operators in their planned programs. If the target program is relatively simple, participants (P2-6) program it without the help of \textit{debug} operator in a short time (2-3 minutes), suggesting the \name API is easy to learn and use.

Our results also demonstrate our API's flexibility, which allows developers to easily explore the impact of various design choices on performance and privacy. Multiple participants demonstrated alternate implementations from the ones we show in \S\ref{sec:productivity}. For example, P4 implemented the productivity tracking hub program in a reversed manner: cropping the body image first and then extracting the pose from this cropped image. This alternative design can potentially be more efficient if a space is mostly unoccupied. 

While \name is easy to learn, our user study identified two barriers.
Multiple participants felt confused about a few common features, and we pointed them to the corresponding unit tests. First, participants confused \textit{retrieve} with \textit{select}, and expected it to keep the original data in its output. Second, the data communicated between operators is an array of \name items rather than an individual item, and this led to some confusion when participants intended to process the faces of multiple persons in one image (see Fig.~\ref{fig:hubprogramhellovisitor}). After the study, we added additional unit tests and documentation to address these misunderstandings. 

Finally, participants offered their opinions on \name APIs unprompted, such as ``much easier to specify data collection behavior through this interface than writing complete code,'' ``Android should also have something similar''.

\subsection{Hub Privacy Features}\label{sec:independentfeatures}

An advantage of \name's chained operators is that their structure and well-known semantics make it fairly easy to analyze and modify app behavior. \new{We developed four built-in features to demonstrate the mechanisms.}

We built a simple static analyzer to \textbf{generate natural language privacy descriptions} based on the manifest automatically, which can support users' decision-making in installing a new \name app. For any manifest, we can describe its data collection behavior using a four-element template: \textit{[trigger], the app sends [content data] to [destination], if [condition].} We derived a set of heuristics to annotate the above properties for each edge in the pipeline based on the operator behavior. For example, the analyzer annotates [\textit{content data}] as \textit{face images} after processing by a cropping face operator. The analyzer annotates the [\textit{condition}] if there is a join operator in the manifest and [\textit{trigger}] based on the \textit{inject} or \textit{push} operator. 
The analyzer stops the annotation when it traverses the whole graph and uses the derived properties to explain the data collection behavior of each network operator. 
Table~\ref{tab:explanations} enumerates explanations generated for apps in \S\ref{sec:casestudies}.



\begin{table}[t]
\renewcommand*\arraystretch{1.9}
	\begin{center}
		\captionof{table}{Auto-generated explanations for 3 case studies using a simple template. We highlight the properties of the template using underlines: \dotuline{trigger}, \underline{content data}, \underline{\underline{destination}} and \dashuline{conditions}.}
		\small 
			\begin{tabular}{ ||c | p{2.8in}  || }
			\hline
			App \# & Generated privacy explanations \\ \hline
			\# 1 & \dotuline{For every week}, the app sends \underline{duration data aggregated} \underline{by content category} to \underline{\underline{www.abc.com}}.\\ 
			\# 2 & \dotuline{When the microphone detects a trigger phrase}, the app sends \underline{anonymized speech audios} to \underline{\underline{www.abc.com}}. \\ 
			\# 3 & \dotuline{For every 30 minutes}, the app sends \underline{extracted poses} to \underline{\underline{www.abc.com}}. \newline
            \dotuline{For every 30 minutes}, the app sends \underline{cropped person} \underline{images} to \underline{\underline{www.abc.com}} if \dashuline{the app cannot recognize poses from the raw image}. \\
			\hline 
		\end{tabular}
		\label{tab:explanations}
	\end{center} 
	\vspace{-0.15in}
\end{table}

\begin{figure}[h]
	\centering
	\includegraphics[width=\linewidth]{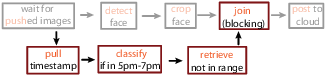}
	\caption{By inserting the subgraph of time checking (highlighted in red boxes), the modified HelloVisitor manifest will not send data outside between 5 pm and 7 pm.
	}
	\label{fig:scheduling}
	\vspace{-0.1in}
\end{figure}

We also implemented a \textbf{time-based scheduling} feature, which can pragmatically modify a \name manifest by inserting a subgraph of operators (highlighted in red boxes in Fig.~\ref{fig:scheduling}), so the app cannot send data outside at certain times.
Imagine that a family does not want their faces captured by the video doorbell when they arrive home, and most family members usually arrive between 5 pm and 7 pm. Based on users’ time specifications, this feature can automatically insert a time-checking branch (i.e., pull->classify->retrieve) to check the time, and merge the two branches using a blocking join operator. In doing so, the data flow can only reach the network operator if the time condition qualifies.

Another feature we built is \textbf{pull rate-limiting}, which allows users to control the frequency at which a smart home app pulls data by modifying the operator properties. Figure~\ref{fig:waterleakfloor} illustrates a water leak detection app, which pulls an image from an existing smart camera every 30 minutes. Using \textit{pull rate limiting}, the runtime can modify the ``interval'' property of the \textit{inject} operator from \textit{30 minutes} to \textit{120 minutes}, making the app only pull images every two hours.

\begin{figure}[h]
	\centering
	\includegraphics[width=\linewidth]{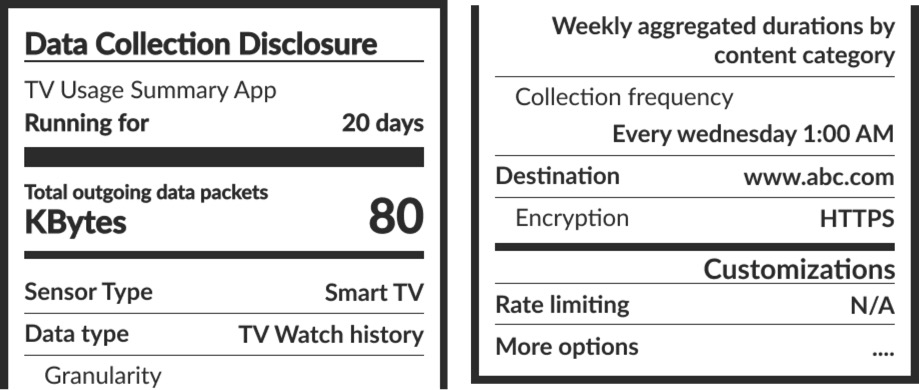}
	\caption{\new{A ``live'' privacy nutrition label generated by the \name runtime automatically.}
	}
	\label{fig:nutritionlabels}
	\vspace{-0.1in}
\end{figure}
\new{Lastly, as a proof of concept, we combined the 3 above features to generate live ``privacy nutrition labels'' that summarizes an app's behaviors (Fig.~\ref{fig:nutritionlabels}).
Apple now requires iOS app developers to fill in a form to create a self-reported ``nutrition label'' for privacy disclosures. However, it can be hard for developers to accurately fill out these forms. 
Furthermore, Apple and third parties cannot easily verify these declared behaviors. In contrast, \name's labels can be auto-generated, thus requiring less effort from developers, and always reflect the actual data practice.}

\section{Related Work}



\vspace*{0.05in}\noindent \textbf{Manifest \& operators}: 
There are many examples of using manifests to create a sandboxed environment to contain untrusted applications (e.g., Janus~\cite{goldberg1996secure}, Android permissions~\cite{AppManif8:online}, MUDs~\cite{WhatisMU82:online}). In contrast to most existing manifests, which often confine applications’ access to system resources (e.g., network, storage, sensors), \name's manifest confines access to data content (e.g., face images, speech audio) in a fine-grained manner. 
Due to the vast number of data granularities, the traditional manifest representation (i.e., raw access enumeration) is no longer feasible. \name addresses this challenge by allowing developers to assemble desired APIs by wiring together a fixed set of operators.

The design of \name operators is inspired by Unix pipes and PrivacyStreams~\cite{privacystreams}. 
PrivacyStreams splits single-pipeline Android data processing into a number of reusable SQL-like operators (e.g., sortby, filter, groupby), which can make data processing more transparent. 
\new{The key innovations of \name are the integrated designs (a manifest, a fixed set of operators, and a trusted runtime with pre-loaded implementations) and the demonstration of using a small and fixed set of operators to support a large number of data pre-processing scenarios.
These two ideas allow \name to offer many features that PrivacyStreams cannot offer, such as OS-level enforcement (i.e., developers can only collect data they claimed in the manifest), additional built-in privacy features, and explicit declarations of fine-grained data granularity (e.g., conditions).}

\noindent \textbf{Privacy-sensitive software architectures}:  
A number of privacy-sensitive architectures for IoT have been proposed as alternatives to conventional cloud-based designs. One example is to process all sensor data at the ``edge'', thereby avoiding sending data to the cloud~\cite{yu2018pinto}. Another approach is to use trusted cloud computing
and perform data minimization through technologies like DIY hosting~\cite{palkar2017diy}, privacy-sensitive machine learning~\cite{chi2018privacy, gilad2016cryptonets,lee2019occlumency,mao2018privacy}, and federated learning techniques~\cite{konevcny2016federated}. However, these approaches often come with tradeoffs such as sacrificing computational efficiency~\cite{lee2019occlumency,li2020federated}, development flexibility~\cite{gilad2016cryptonets,li2020federated}, or service quality~\cite{valerio2016accuracy}.

In contrast, \name offers a hybrid approach that does some pre-processing locally on the hub while also allowing developers to use cloud services in a manner that they are accustomed to~\cite{10.1007/978-3-642-30973-1_32}. \name's hub is similar in spirit to cloudlets~\cite{satyanarayanan2009case}, but the key difference is that the computation running on the hub is structured and enforced using the operator-based manifest. \name is also inspired by past smart home hub/gateway/firewall projects~\cite{chi2019pfirewall,demetriou2017hanguard,ko2018deadbolt,DBLP:journals/corr/abs-2105-05162,zhang2021capture}.
\name has two major differences. First, \name enables data minimization at the data content level (e.g., only sending face images), while existing projects can only block individual outgoing network requests. Second, \name requires developers to explicitly declare the data collection behaviors, facilitating auditing and enforcement.

\vspace*{0.05in}\noindent \textbf{Privacy awareness and control}: A complementary approach to privacy is better mechanisms for notice and choice~\cite{agarwal2013protectmyprivacy,dixon2012operating,huang2020iot,liu2011analyzing,khandelwal2021prisec}, e.g. at runtime for mobile apps~\cite{Requesti97:online, nauman2010apex}, often extended to smart home contexts~\cite{dixon2012operating}. In the case of IoT privacy, merely allowing or denying sensor access is insufficient and this all-or-nothing access control either exposes sensitive data or breaks the app functionality. In response to this, recent efforts offer finer-grained control and transparency, particularly on the fidelity of the data~\cite{aditya2016pic,6547120,raval2016you,wang2017scalable}. 



However, privacy support in these systems is often built on an individual basis with no common structure, imposing challenges for both app development and user privacy management. 
Third-party auditors also cannot easily verify if these features 
work as claimed~\cite{10.1145/3427228.3427277,Teardown30:online}.
\name addresses these needs through a novel privacy architecture, which can enable transparent and enforceable data collection, and offer centrally manageable hub privacy features.

\section{Discussion}

\vspace*{0.05in}\noindent \textbf{\new{Architecture adoption}}: 
\new{Many requirements for the \name hub are well aligned with the roles of recent commercial ``hub'' products, which may facilitate adoption. For example, many hubs (e.g., Philips Hue Smart Hub) serve as a central in-home gateway to connect devices, mediating access between the internal devices and the Internet. In addition, most hubs (e.g., Nest Hub Pro) have a moderate amount of computing power to pre-process out-going data and offer a centralized hub user interface. A few hubs (e.g., Samsung SmartThings Hub) even behave like an early app store.}

\vspace*{0.05in}\noindent\textbf{\new{Alternative implementations}}: \new{\name's manifest is a new program representation that offers more flexibility than all-or-nothing access, while being more structural and verifiable than arbitrary code (e.g., Java). Future work can also implement the manifest in other flow-based programming frameworks (e.g., NoFlo, Pyperator)~\cite{samuella28:online}. We chose Node-Red since it is popular in the home automation hobbyist community, provides many open-source device drivers, and has a mature user interface.} 


\vspace*{0.05in}\noindent \textbf{\new{Design pattern adoption}}: 
\new{The design of \name can be generalized as a reusable design pattern for cases where first- and third-parties are trying to access sensitive data, e.g., browser plugins, calendar APIs, and smartphones~\cite{10.1145/3081333.3081334}. Our core ideas of a fixed set of operators, a text-based manifest where all outgoing data flows must be declared, and a trusted computing platform with pre-loaded implementations can thus be useful in these cases. For example, Google Calendar only allows users to grant all-or-nothing access to third-party developers. However, most apps (e.g., Zoom) do not need full access. A future calendar API might offer a set of common operators instead and allow developers to program their access in \name-like manifests. This design pattern can make data flows transparent, enforce data transformations, and allow third-parties to build independent privacy features. 
}

\vspace*{0.05in}\noindent \textbf{\new{Role of users, developers, auditors in determining privacy-utility trade-offs}}: 
\new{Peekaboo has the potential to disincentivize overcollection of data. We expect \name manifests to be public, making it possible for app stores and third-party auditors (e.g., Consumer Reports) to analyze manifests programmatically at scale. Users can also see if the required data granularity make sense, and flag items in a review if they do not, block certain outgoing data, or choose not to install an app. Altogether, this kind of transparency has the potential for nudging developers to collect less data.}

\vspace*{0.05in}\noindent \new{\textbf{Hosting proprietary algorithms}: Some of the best implementations of inference mechanisms such as keyword spotting, keypoint tracking, and biometric authentication can be proprietary. 
Platform/hub builders may implement these algorithms inside their hub drivers in the future, or hardware developers can provide the functionality directly on the edge devices.}


\vspace*{0.05in}\noindent \textbf{Extensibility}: 
\name assumes a fixed set of operators and a stable data model to support its data pre-processing pipelines. 
Although the taxonomy in Fig.~\ref{fig:actiontaxonomy} may not be complete, we anticipate that the list and semantics of operators will converge quickly and remain stable for years. 
\new{Further, platform builders can extend the operator options by expanding supported data transformations (e.g., removing all audible frequencies from the audio~\cite{10.1145/3411764.3445169}) and adding new pre-trained models.
Platform builders can also develop new drivers to support more devices.}


\new{We expect the runtime to be updated over time, analogous to getting a new version of Linux or Java. Also, similar to Android's manifest, a \name manifest will need to specify a minimum required runtime version. We do not expect the pre-loaded operators to grow into a large library of data transformation functions. Instead, we believe a few simple and common data transformations (e.g., tabular data aggregation, image cropping/blurring) can cover many common scenarios and go a long way towards improving privacy.}


\vspace*{0.05in}\noindent \new{\textbf{Benefits of \name}: \name has three important advantages over building data minimization features individually. (1) \textbf{Transparency}. Individually built data minimization practices are black boxes to outsiders. Indeed, even if developers open-source their products or allow a third-party auditor to access their codebases, inspecting the actual data collection behavior is difficult. (2) \textbf{Ease of development}. Building data minimization algorithms and user interfaces for privacy requires significant effort. By authoring a \name\ manifest, \name developers can leverage many built-in features for free. (3) \textbf{Centralized privacy management}. If all developers build management interfaces individually, users would have to deal with potential inconsistencies between these user interfaces and their different semantics. In contrast, Peekaboo can offer centralized, fine-grained features across devices.}


\section{Future work \& Limitations }\label{sec:futurework}

\vspace*{0.05in}\noindent \textbf{End-to-end encryption}: \name currently requires two separate encrypted  connections (i.e., devices-to-hub and hub-to-servers) for its operation rather than end-to-end encryption from device to server directly. While SSL proxy mechanisms (e..g, ~\cite{TLSinspe82:online, SSLForwa90:online,7927897, PolarPro55:online}) may provide a way to support end-to-end encrypted connections with the ability to verify what data is being sent, it is not clear whether they can support \name's operators that transform data. This is a current limitation, and we defer this to future work.


\vspace*{0.05in}\noindent \textbf{More hub features}: Beyond the \new{four} hub privacy features introduced in \S\ref{sec:independentfeatures}, further work may explore many other hub privacy features by analyzing and rewriting the manifest program. 
\new{For example, the hub may aggregate the installed manifests, 
make privacy nutrition labels interactive, 
enable centralized privacy dashboards}, and allow users to query  
\new{what/when/how} data flows out. The hub can also potentially allow users to apply global filters (e.g., blocking outgoing face images) across manifests, e.g. to make guests more comfortable with cameras around the house.

\vspace*{0.05in}\noindent \new{\textbf{Manifests for other communication patterns}: 
\name focuses on whitelisting edge-to-cloud data flows to improve privacy. A promising research direction is to generalize the manifest for other communication patterns, such as device-to-device, cloud-to-device, or even physical actuation. Ideally, we may have a set of Peekaboo-like manifests for each IoT device, whitelisting its interactions with the rest of the world in a common, structured, useful, and understandable way. Such an infrastructure can help users establish a correct mental model of device behaviors, allow the hub to offer built-in protection for the physical events (e.g., a bread toaster cannot run for more than an hour), and ease software development.}

\vspace*{0.05in}\noindent \textbf{Unpredictable hub program rewriting}. In this paper, we demonstrate the feasibility of several features that rewrite the hub program written by developers. However, such program rewriting may break the original functionality if users misuse the feature. For example, a user may choose to blur their faces in an unlocking app, which makes her unrecognizable to the unlocking app. 

\vspace{-0.2cm}
\section{Conclusion}
This paper introduces \name, a new IoT app development framework to help developers build privacy-sensitive smart home apps. \name offers a hybrid architecture, where a local \textbf{user-controlled hub} pre-processes smart home data in a structured manner before relaying it to external cloud servers. In designing Peekaboo, we propose three key ideas: (1) factoring out repetitive data pre-processing tasks from the cloud side onto a user-controlled hub; (2) supporting the implementation of these tasks through a fixed set of open-source, reusable, and chainable \textbf{operators}; (3) describing the data pre-processing pipelines in a text-based \textbf{manifest} file, which only stores the operators' specifications, but not the operators' actual implementation.

\section*{Acknowledgement}

This research was supported in part by the National Science Foundation under Grant No. CNS-1801472, CNS-1837607, and CNS-2007786, Cisco, Intel, Infineon, and Air Force Research Laboratory under agreement number FA8750-15-2-0281. We thank the anonymous reviewers for their constructive feedback.

\balance
\bibliographystyle{plain}
\bibliography{peekaboo}
\onecolumn
\appendix
\section*{A. Design fiction interview apparatus}\label{sec:designfiction}
We conducted design fiction interviews~\cite{dunne2013speculative} with 7 participants (3 female, 4 male) across different age groups (min=21, max=60, avg=32) to broaden our set of smart home use cases.
Design fiction is a speculative design method, where participants are asked to sketch diegetic prototypes in an imaginary story world.
All participants had interacted with multiple smart home products before. For the interview, we presented each participant with a hypothetical scenario\footnote{Adapted from~\cite{dixon2012operating}} and a floor plan of a home, asking each participant to play the role of different family members and brainstorm different smart home functionality they desired. We also asked participants to 
focus on the role’s needs, rather than feasibility of sensing or hardware availability.

\textbf{The hypothetical scenario}: \textit{Imagine a family of five members living in a single-family house. Jeff and Amy are husband and wife respectively. Dave and Rob are their eight-year and seven-month-old kids. Jeff’s mother, Jen, helps look after the baby, and Jeff’s brother Sam occasionally visits the house. The family plans to purchase a set of devices to enable different smart home applications.}

\textbf{The role playing questions}: \textit{If you are [Jeff|Amy|...], what smart home usages do you want in the [basement|living room|...]?}

\begin{figure}[h!]
	\centering
	\includegraphics[width=0.59\columnwidth]{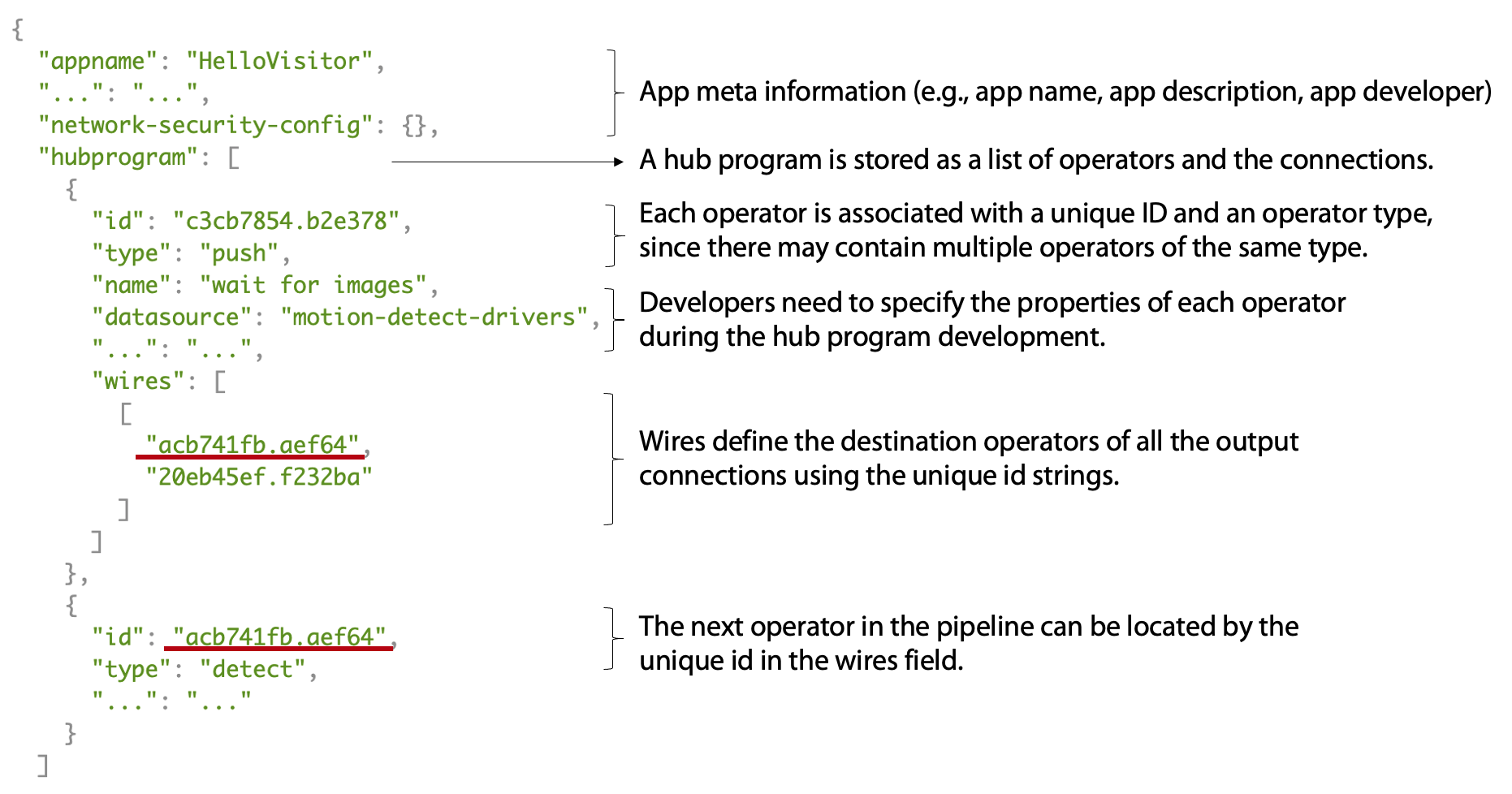}
	\caption{The manifest file of a hub program is stored in JSON format. The manifest file contains three types of information: app meta information, security configuration, and the hub program presentation.}
	\label{fig:appmanifest}
	\vspace*{-0.1in}
\end{figure}

\begin{table}[h!]
	\renewcommand{\arraystretch}{1.2}
	\begin{center}
		\captionof{table}{Supported data transformation across data types.}   
        \small
		\begin{tabular}{ || c | p{14mm}  | p{120mm} || }
			\hline	\hline	
			Data type & Operator & Supported transformations \\
			\hline	\hline	
			\multirow{2}{*}{Video}
             & Extract & Extracting heart rates using~\cite{balakrishnan2013detecting} \\ \cline{2-3}
                  & Retrieve & Keeping inference results but removing the video data \\ \hline
            \multirow{2}{*}{Image} & Detect & Detecting the bounding box of 90 Common objects (Microsoft COCO~\cite{lin2014microsoft}) and faces, Detecting the segmentation areas of 20 common object segmentation (PASCAL VOC2012~\cite{chen2017deeplab})\\ \cline{2-3}
            & Extract & Extracting the brightness, body poses~\cite{Poseesti0:online} \\ \cline{2-3}
            & Noisify & Blurring an image \\ \cline{2-3}
            & Select & Cropping an image based on bounding boxes or a segmentation areas \\ \cline{2-3}
            & Retrieve & Keeping inference results but removing the  image data\\ \hline
            \multirow{2}{*}{Audio} & Detect & Detecting voice activity windows (i.e., starting time and ending time)  \\ \cline{2-3}
            & Classify & Recognizing 632 audio events (AudioSet~\cite{audioontoloy}) \\ \cline{2-3}
            & Extract & Extracting frequency spectrum, speech text \\ \cline{2-3}
            & Noisify & Injecting a configurable random variation to the pitch and tempo. \\ \cline{2-3}
            & Retrieve & Keeping inference results but removing the  audio data\\\hline
            \multirow{2}{*}{Tabular} & Select & Selecting a column with an optional where clause\\ \cline{2-3}
             & Aggregate & Aggregating (i.e., sum, count, average) the tabular entries by one field and projecting the output to a designated field.  \\ \hline
            \multirow{2}{*}{Scalar} & Classify & Comparing a value with a threshold \\ \cline{2-3}
            & Aggregate & Computing the sum, count, average of matched scalar items.\\ \cline{2-3}
            
            & Retrieve & Keeping inference results but removing the  original scalar data\\
			\hline	\hline	
		\end{tabular}
		\label{tab:datatransformations}
	\end{center}
\end{table}

%




\begin{table}[h!]
	\renewcommand{\arraystretch}{1.25}
	\begin{center}
		\captionof{table}{A summary of collected smart home use cases drawn from the research literature and design fiction interviews we conducted (\S\ref{sec:feasibility}). We obtained the initial list of use cases from the design fiction interview (Appendix \S\ref{sec:designfiction}), which asked participants to brainstorm use cases in each room using a single-family house floor plan. We then augmented this list by enumerating different sensors supporting these applications and potential data collections that developers may need. 
This list is not exhaustive, but our goal is to have a large sample covering many common smart home use cases. }   
        \small
		\begin{tabular}{ || c | p{105mm}  | p{50mm} || }
			\hline	\hline	
			ID & Description (Location) & Relevant sensors \\
			\hline	\hline	
			\#1 & Water leak detection on the floor  (Basement, Kitchen, Garden, Bathroom) & Camera, humidity, microphone\\
			\hline	
			\#2 & Home inventory tracking  (Storage room, Closet) & Camera, RFID \\
			\hline	
			\#3 & Toxic Gas Alarm (Basement, Kitchen) & Specialized sensors  \\
			\hline
			\#4 & Temperature tracking across rooms (The whole house) & Themometer\\
			\hline
			\#5 & Garage usage detection (Garage, garden) & Camera, occupancy, microphone \\
			\hline
			\#6 & Street parking spots detection (Driveway) & Camera, occupancy, radar\\
			\hline
			\#7 & Home arrival recording/prediction (Entrance doors, Garage) & Camera, RFID \\
			\hline
			\#8 & Garden irrigation tracking (Garden, driveway) & Camera, humidity \\
			\hline
			\#9 & Soil health tracking (Garden, driveway) & Specialized sensors\\
			\hline
			\#10 & Bath room activity (e.g., toileting for elderly) tracking (Bathrooms) & Humidity, occupancy, camera, pressure \\
			\hline
			\#11 & Smart Speaker \& Voice control TV (Living rooms, Bedrooms) & Microphone, proximity \\
			\hline
			\#12 & Room occupancy statistics (The whole house) & Camera, occupancy, microphone \\
			\hline
			\#13 & Automatic temperature control based on the occupancy/identity (The whole house) & Camera, occupancy, microphone \\
			\hline
			\#14 & Office productivity tracking (Office room) & Camera \\
			\hline
			\#15 & Personalizing welcome message for visitors (Entrance doors, garage) & Camera \\
			\hline 
			\#16 & Package delivery detection (Entrance doors) & Camera \\
			\hline
			\#17 & Fall detection for the elderly (The whole house) & Camera, microphone \\
			\hline
			\#18 & Automatic photography~\cite{GoogleAI96:online} (Living room, Kitchen, Garden) & Camera  \\
			\hline
			\#19 & Automatic lighting based on occupancy and light intensity (The whole house) & Occupancy, light sensors, camera \\
			\hline
			\#20 & Wanted criminal search on the street (Entrance doors) & Camera \\
			\hline
			\#21 & Detecting strangers when no one is at home (The whole house) & Camera \\
			\hline
			\#22 & Ice detection (Entrance stairs, driveway) & Camera, specialized sensors\\
			\hline
			\#23 & Sunshine tracking for plants (Terrace, garden) & Camera, light sensors \\
			\hline
			\#24 & Detecting messiness of the home and calling a cleaning service (Living room, kitchen, closet) & Camera \\
			\hline
			\#25 & Smart cooking (Kitchen) & Smart appliances \\
			\hline
			\#26 & Freezer ice cleaning reminder (Kitchen) & Smart appliances \\
			\hline
			\#27 & Appliances electricity consumption statistics (The whole house) & Smart appliances and plugs \\
			\hline
			\#28 & Sleep tracking and sleep quality measurement (Bedrooms)  &   Camera, microphones, pressure sensors  \\
			\hline
			\#29 & Smart toilet recognizes butt and analyzes poop for diseases~\cite{park2020mountable} (Toilets) & Camera, pressure sensors  \\
			\hline
			\#30 & Detecting baby crying (The whole house) & Camera, microphone \\
			\hline
			\#31 & Laundry service reminder (Closet, the whole house) & Camera, RFID  \\
			\hline
			\#32& Ubiquitous bio-metric measurement (e.g., height, heart rate) ( the whole house) & Camera, heart rate sensor, Wi-Fi \\
			\hline
			\#33 & Pet barking detection (The whole house) & Camera, microphone \\
			\hline
			\#34&  Smart stylists that locates the clothes (Closet) & Camera, RFID  \\
			\hline
			\#35&  Water activity detection (The whole house) & Specialized sensors  \\
			\hline
			\#36&  Measuring lung function using a microphone & Microphone \\\hline
			\#37&  Acoustic ranging for games and cross-device interaction & Microphone \\
			\hline	\hline	
		\end{tabular}
		\label{tab:exampleiotapps}
	\end{center}
\end{table}


\begin{table*}[h!]
	\renewcommand{\arraystretch}{1.15}
	\begin{center}
		\captionof{table}{What our team felt were reasonable outgoing data granularities for different smart home scenarios. We cluster these scenarios by their input and the output. If the output is Original, it implies no pre-processing functions can be done on the hub for these scenarios. The "(\#scenario)" is the number of scenarios for each corresponding category, and the numbers in the "scenarios" column refer to the scenario id in Table ~\ref{tab:exampleiotapps}. Since a use case may have multiple reasonable app designs, the use case may be counted more than once.}
        \small
		\begin{tabular}{ || p{18mm} | p{28mm}  | p{75mm} | p{28mm} || }
			\hline	\hline	
			\makecell{ Data type \\ (\# scenarios) }  &  \makecell{Pre-processed output  \\ (\# scenarios)}& Example usage & Scenarios  \\  \hline \hline
			\multirow{2}{18mm}{{Image/Video (27)}} & \textbf{Original}    &    & \\
			& - raw video \hfill 1 & Enabling online security camera album & {  \#21 } \\
			& - raw image   \hfill 3  & Enabling automatic photography (e.g. Google Clips~\cite{GoogleAI40:online})  & { \#18}  \\ 
			& 	  & Recognizing special activities (e.g., garden irrigation)  &  {  \#8, \#29}    \\ 
			& \textbf{Partial original}    &    & \\ 
			& - only person/face  \hfill 10  &  Recognizing human relevant activities (e.g., fall, sleep)  & {  \#7, \#12, \#13, \#14, \#17, \#20,  \#21,  \#28,  \#30,  \#32} \\ 
			&{ - excluding person   \hfill 7 }& Determining home messiness  &  {\scriptsize \#1, \#2, \#5, \#19, \#23, \#31,   \#34} \\ 
			&{  - particular objects   \hfill 3 } & Detecting certain objects (e.g., package) &  {  \#16, \#22,  \#33} \\ 
			& \textbf{Derived}   &    & \\ 
			& {  - identity  \hfill 2 }&  Personalizing the welcome message  & {  \#7, \#15 }\\ 
			& {  - pose \hfill 2 } &  Quantifying work activities  & {  \#14, \#17  }\\ 
			& { - light intensity  \hfill 2 }&  Determining the lightness of the environment  & {   \#19, \#23, }\\ 
			\hline
			\multirow{2}{*}{{Audio (9)}} 	& \textbf{Original}    &    & \\ 
			& - raw audio \hfill 2 & Supporting signal processing at per-frame level  & {  \#36, \#37}  \\
			&                   \hfill 1 & Supporting phone call, audio diary  & {  \#11} \\
			& \textbf{Partial original}    &    & \\ 
			& - voice audio  \hfill 1 & Supporting speech interaction   & {  \#11 } \\ 
			& - activity sound  \hfill 3  &  Detecting garage events  & {  \#5, \#28, \#30} \\ 
			& \textbf{Derived}    &    & \\ 
			& - speech text  \hfill 1 &  Recognizing command intents   &  {   \#11}\\ 
			& - FFT/MFCC   \hfill 2  & Detecting occupancy  & {   \#12, \#13} \\ 
			& - {  audio events}   \hfill 3 &  Sending notification to owners when the dog barks   & {    \#12, \#13, \#33} \\ 
			\hline
			\multirow{3}{18mm}{{Tabular (e.g., Channel state information) (5)}}  & \textbf{Original}    &    & \\ 
			& - complete CSI \hfill 1  & Enabling fine-grained sensing tasks using the Wi-Fi signals  &  {   \#32, \#34} \\
			& \textbf{Partial original}    &  &  \\ 
			& - only UUID \hfill 3  & Tracking the food/cloth inventory & {    \#2, \#7, \#30} \\
			& \textbf{Derived}    &    & \\ 
			& - distance/positions  \hfill 1 & Helping users find the item in home & {   \#34 } \\\hline
		\multirow{4}{18mm}{Scalar (e.g. temperature, location, moisture) (13/37)}
			& \textbf{Original}    &    & \\ 
			&  - humidity \hfill 2 &   Detecting water leakages using precise humidity records  & {   \#1 , \#8}   \\ 
		    & - water pressure \hfill 1  &  Detecting in-home activity using precise humidity records & {   \#35} \\ 
		    & - misc  \hfill 8 & Most scalar values are relatively safe to share & {  \#3, \#4, \#5, \#6, \#22, \#26, \#27, \#29}  \\
			& \textbf{Partial original}    &    & \\ 
			& -  coarse humidity  \hfill 1 &  Detecting showering events in the bathroom  & {   \#10} \\
			& \textbf{Derived}    &    & \\
			& - out-of-town status  \hfill 1 & Detecting if are users out-of-town to adjust the AC temperature &   {   \#13}\\
			\hline
			\hline
		\end{tabular}
		\label{tab:exampledatalevels1}
	\end{center}
\end{table*}

\end{document}